\documentclass[aps,longbibliography,superscriptaddress,showpacs,10pt,twocolumn]{revtex4-1}
\newif\ifletter\letterfalse 
\usepackage{amsmath}
\usepackage{graphicx}
\usepackage[hidelinks]{hyperref}
\usepackage[T1]{fontenc}
\usepackage[utf8]{inputenc}
\usepackage{amssymb}
\usepackage{amsthm}
\usepackage{bbm}
\usepackage{mathtools}

\newtheorem{thm}{Theorem}

\newtheorem{lem}[thm]{Lemma}
\newtheorem{prop}[thm]{Proposition}

\def\Qref#1{\splitref#1@}
\def\splitref#1:#2@{\thing#1@\;\ref{#1:#2}}
\def\thing#1#2@{\ifx#1sSect.\else\ifx#1fFig.\else\ifx#1TTable\else\ifx#1tThm.\else\ifx#1dDef.\else\ifx#1lLem.\else\ifx#1cCor.\else\ifx#1pProp.\else\ifx#1PProperty\fi\fi\fi\fi\fi\fi\fi\fi\fi}

\ifletter

\def\llsection#1{}

\def\coloronline{Color online.\ }

\else
 \let\llsection\section

\def\coloronline{}
\fi

\def\tr{\operatorname{tr}}
\def\idty{{\leavevmode\rm 1\mkern -5.4mu I}} 

\def\Rl{{\mathbb R}}\def\Cx{{\mathbb C}}
\def\Nl{{\mathbb N}}

\def\norm #1{\Vert #1\Vert}

\def\order#1{{\mathbf o}(#1)}

\def\braket#1#2{\langle #1,#2\rangle}

\def\brAAket#1#2#3{\langle#1\vert#2\vert#3\rangle}
\def\bra#1{{\langle#1\vert}}
\def\ket #1{\vert#1\rangle}
\def\ketbra #1#2{{\vert#1\rangle\langle#2\vert}}
\def\kettbra#1{\ketbra{#1}{#1}}

\def\tr{\mathop{\rm tr}\nolimits}
\def\abs#1{\vert#1\vert}

\let\veps\varepsilon

\def\re{\Re e}
\def\im{\Im m}

\def\intd#1{\int\mskip-7mu d#1\ }  

\def\BB{{\mathcal B}}\def\HH{{\mathcal H}}
\def\MM{{\mathcal M}}

\def\minivec(#1,#2){{\textstyle\begin{pmatrix}#1\\#2\end{pmatrix}}}

\def\wig{{\mathcal W}} 
\def\cdt{{\cdot}} 
\def\jnr{{\mathcal R}} 
\def\sing{{\mathcal S}} 

\def\twirl{{\mathbb P}_R} 
\def\dihed#1{{\mathsf D}_{#1}}

\begin{document}
\title{The Wigner distribution of $n$ arbitrary observables}

\author{Ren\'e Schwonnek}
	\email{Rene.S@nus.edu.sg}
	\affiliation{Institut f\"ur Theoretische Physik, Leibniz Universit\"at, Hannover, Germany}
	\affiliation{Department of Electrical and Computer Engineering, National University of Singapore, Singapore}
	\affiliation{Centre for Quantum Technologies, National University of Singapore, Singapore}
\author{Reinhard F. Werner}
	\email{Reinhard.Werner@itp.uni-hannover.de}
\affiliation{Institut f\"ur Theoretische Physik, Leibniz Universit\"at, Hannover, Germany}

\date{\today}
\begin{abstract}
We study a generalization of the Wigner function to arbitrary tuples of hermitian operators.
We show that for any collection of hermitian operators $A_1,\ldots,A_n$, and any quantum state there is a unique joint distribution on $\Rl^n$, with the property that the marginals of all linear combinations of the $A_k$ coincide with their quantum counterpart. In other words, we consider the inverse Radon transform of the exact quantum probability distributions of all linear combinations. We call it the Wigner distribution, because for position and momentum this property defines the standard Wigner function.
We discuss the application to finite dimensional systems, establish many basic properties and illustrate these by examples.  The properties include the support, the location of singularities, positivity, the behavior under symmetry groups, and informational completeness.\vspace{0.1cm}
\end{abstract}

\ifletter\fi
\maketitle
\vskip 1cm

\section{Introduction}
It is well known that non-commuting hermitian operators do not admit a joint measurement, i.e., their spectral measures cannot be written as the marginals of some other positive operator valued measure. That is, there is no way to linearly assign to each quantum state (density operator) a positive joint distribution, whose marginals are the standard probability distributions of the given operators.  The positivity in this statement is crucial, and Wigner's quasiprobability function \cite{Wigner} is the paradigm of a non-positive operator valued measure, whose marginals are the momentum and position observables, respectively.

Wigner functions have been introduced in 1932 \cite{Wigner} originally for treating the classical limit. Since then they have become a key tool for understanding systems with canonical degrees of freedom like position/momentum, or field quadratures in quantum optics. They have been particularly helpful in providing a visualisation of quantum states. While listing all the matrix elements of the density matrix is a correct way to completely characterize a state, it is hard to present this information in a way that would help to gain an useful insight. The Wigner function of a state contains the same information, but combines it with a localization in phase space, so that some properties, like negativity of the Wigner function, can be associated with particular regions of phase space, carrying direct suggestions as to what experiments might see these effects. Accordingly, there is a rich literature on Wigner functions \cite{OConnell1983,Schleich,Hillery,Cohen,Hudson,Grossmann,Kenfack,Ferrie,Wer88,Hornberger,Gosson}. 
\begin{figure}[ht]
	\vskip0.4cm
	\includegraphics[width=0.90\columnwidth]{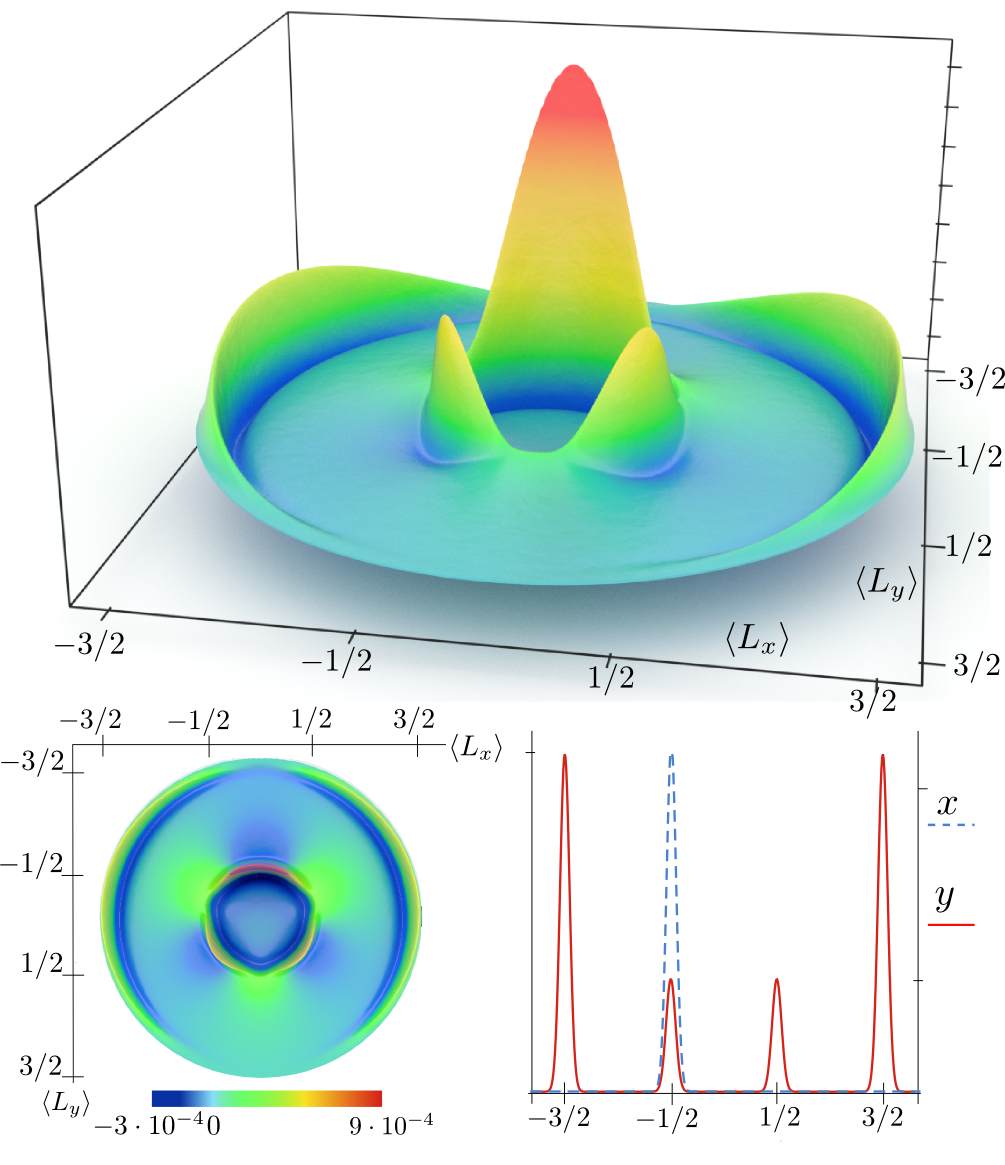}
	\caption{\coloronline Wigner function of an angular momentum measurement in z- and x direction on a spin 3/2-system with
		$\psi= |m,j\rangle=|-1/2,3/2\rangle$. Gaussian regularization with $\veps=0.001$.
		Lower left: Level sets encoded with blue$\to$negative, red$\to$positive.
		Lower right: Marginal distributions in $z$-direction (red) and $x$-direction (blue).}
	\label{fig:3rand}\vskip0.2cm
\end{figure} 

There are various generalizations of the Wigner function focusing on the group theoretical structure from the position/momentum case \cite{Ali2000,Klimov,Wootters,Gross,Kastrup,Kastrupp}.  Here, the Wigner function typically is a function on the underlying group. Even though generalizations of this kind are useful in their specific context, they do not necessarily provide a satisfactory definition if we consider an arbitrary set of observables,
since, in this case there is generally no symmetry that would justify the choice of a particular group.   

In contrast, the aim of this note is to explore another generalization that is based generalizing algebraic properties of the Wigner function, for a general set of observables $A_1,\dots, A_n$. 

In Detail our generalization is directly based on the well-known fact, that the Wigner function not only gives correct marginals for canonical momentum $P$ and position $Q$, but also for all real linear combinations of these operators. More precisely,  for all coefficients $\alpha,\beta\in\mathbb{R}$ and every $k$-th moment and we have the identity
\begin{align}
\left\langle \left(\alpha P +\beta Q \right)^k\right\rangle_\rho=\Bigl\langle \left(\alpha P +\beta Q \right)^n\Bigr\rangle_{\mathcal W_\rho (p,q)}\label{eq:moments}
\end{align}
In the above, the l.h.s  is evaluated 'quantum mechanically', this is on the level of operators $(P,Q)$ and for a given quantum state $\rho$. The r.h.s., however, is evaluated for 'quasi random variables' distributed by an according quasi probability distribution $\mathcal{W}_\rho(p,q)$ with support on $\mathbb{R}^2$.

As we will show, replacing the pair $(P,Q)$ by $n$ arbitrary hermitian operators $(A_1,\ldots,A_n)$  in \eqref{eq:moments} singles out a unique distribution of $n$ variables, which we call the Wigner distribution. Apart from the operators $A_k$ and the state there are no arbitrary choices or parameters in this definition. 

To our knowledge an equivalent definition has only been suggested within a work of Cohen and Scully from the 80s \cite{CohScu} (see also \cite{ScuCoh}, \cite[Ch.~17]{Cohen}). However, since then it apparently has never been worked out in any detail.  We will extend this by general results, a numerical method for obtaining graphs at any desired resolution, and further examples. We hope to thereby convey an impression of the kind of quantum structures that can be seen in this representation, and what intuitions might be based on it. A basic example, demonstrating that even in simple cases the Wigner function may display a lot of structure, is shown in Fig.~\ref{fig:3rand}.


Regarded from a more general perspective, what we propose is merely the application of classical tomography to a set of quantum mechanical data: Suppose we know the collection of probability distributions of the linear combinations $\sum_k\xi_kA_k$ in the state $\rho$. This is exactly the sort of information available from low dimensional projections of higher dimensional structures as in X-ray image analysis. The reconstruction method of choice in that case is the inverse Radon transform \cite{Radon}, and we merely apply this  transform to the quantum mechanical data.  

We know that due to their quantum mechanical origin, the distributions do {\it not} come from a joint probability. The result is nevertheless unique, although it lacks positivity. It  inherits its operational meaning from the marginals, but is especially sensitive to deviations from a classical setup: Both negativity and singularities attest to non-classical features, although, as in the case of the standard Wigner function these indicators are not very direct. For the standard Wigner function a body of interpretational knowledge has been accumulated, but for its generalization we can only make a beginning.

\newpage
\section{Definition}
Let $A_1,\ldots,A_n$ be bounded hermitian operators on a Hilbert space.
We typically take the Hilbert space dimension $d=\dim\HH$ to be finite, as well.
Linear combinations with real coefficients $\xi=(\xi_1,\ldots,\xi_n)$ are written as $\xi\cdot A=\sum_k\xi_k A_k$.
Such an operator tuple will be assumed to be fixed throughout the paper, as well as a density operator $\rho$.

In the following we will give three different but equivalent definitions of the Wigner distribution

\subsection{by the marginal property}
The {\it Wigner distribution} is a distribution (generalized function) satisfying the equation
\begin{equation}\label{WigDef}
\int\!\!d^na\ \wig_\rho(a_1,\ldots,a_n)\ f(\xi\cdt a)= \tr\rho f(\xi\cdt A),
\end{equation}
for every bounded infinitely differentiable function $f{:}\Rl\to\Cx$. If $\wig_\rho$ were a probability density this would just equate the classical expectation of some function of the random variable $\xi\cdt a$ with the quantum expectation of the same function, applied to the operator $\xi\cdt A$.  However, $\wig_\rho$ is usually not a probability density, not just because it may be negative, but because it may be a generalized function/distribution. From this definition it is not immediately clear that such a distribution exists and is even unique. However, this will be clear from the next one.

\subsection{by Fourier transform}\label{sec:FourierDef}
When we take $f(t)=\exp(it)$ in \eqref{WigDef} we directly get an expression for the Fourier transform $\widehat\wig_\rho$ of $\wig_\rho$, namely
\begin{equation}\label{FouWig}
\widehat\wig_\rho(\xi)=\tr\rho e^{i\xi\cdt A}.
\end{equation}
This fixes $\widehat\wig_\rho$ uniquely as a distribution, which we consider as a tempered distribution (i.e., with the Schwartz functions for a test function space). The reverse Fourier transform, formally
\begin{equation}\label{FouFouWig}
\wig_\rho(a)=\frac1{(2\pi)^n}\int d\xi\ e^{-i\xi\cdot a}\widehat\wig_\rho(\xi)
\end{equation}
is then also a tempered distribution, which we denote by $\wig_\rho$. We will establish below (\Qref{prop:supp}) that this distribution has compact support, which makes it possible to integrate it with differentiable, but unbounded ``test functions''.

It follows immediately from the definition and the observation that
$\overline {\widehat\wig_\rho(\xi)}=\widehat\wig_\rho(-\xi)$ that $\wig_\rho$ is real-valued.

\subsection{by ordered moments}
Let $r=(r_1,\ldots,r_n)\in\Nl^n$, and $R=\sum_{k=1}^nr_k$. The we define the {\it Weyl-ordered moment} of order $r$ as
\begin{equation}\label{Weylorder}
M_W^r(A):=\frac1{R!}\sum_\pi A_{i_{\pi1}} A_{i_{\pi2}}\cdots A_{i_{\pi R}}
\end{equation}
where $\pi$ runs over all permutations of $\{1,\ldots,R\}$, where the labels $i_1,\ldots,i_R\in\{1,\ldots n\}$ are chosen so that when $\pi$ is the identity the product is simply $A_1^{r_1}A_2^{r_2}\cdots A_n^{r_n}$.
That is, each product contains $r_1$ factors equal to $A_1$, etc, and $M_W$ is the average over all such products. This form of the definition follows a pattern that can be used also for other for ordered moments\cite{Taoblog1,CohenBook2}, like the ``normal'' or ``Wick'' ordering where the average contains only the term with $\pi$ the identity, or the ``antinormal'' with only the reversing permutation $\pi k=R-k+1$. All these moments coincide with the ordinary ones when the $A_k$ commute. The importance of the Weyl ordering lies in the non-commutative multinomial theorem: For any coefficients $\xi_1,\ldots,\xi_n$ one has
\begin{equation}\label{multinomial}
\sum_r\textstyle{R\choose r} \xi_1^{r_1}\cdots \xi_n^{r_n}M_W^r(A)=(\xi\cdt A)^R.
\end{equation}
Now consider any polynomial $f(a_1,\ldots,a_n)$, expand, and replace the monomial $a_1^{r_1}a_2^{r_2}\cdots a_n^{r_n}$ by $M_W^r(A)$. We denote the resulting operator by $\wig^*_f$, the Weyl-ordered operator version of $f(A)$.  Then
\begin{equation}\label{WigDefmom}
\int\!\!d^n\,a\ \wig_\rho(a)\ f(a)= \tr\rho \wig^*_f.
\end{equation}
(This follows directly from \eqref{WigDef} for the special case $f(a)=(\xi\cdt a)^R$, then by expansion for monomials, and by linear comnbination for all polynomials).
Note that we have defined $\wig^*_f$ directly, so this could be read as a defining condition for $\wig_\rho$. On the other hand, when $\wig_\rho$ is already defined via Fourier transform, (and we know it has compact support, see below) we can turn this around and read \eqref{WigDefmom} as the definition of the quantization map $f\mapsto \wig^*_f$. The range of this map is the span of the Weyl-ordered moments, which we denote by $\MM_W$.

\subsection{Graphical representation}
It may seem that a proper distribution is not suitable for visualization, because  $\wig_\rho(a)$ is not even defined pointwise for every $a$. However, as for all function plots, one needs to specify a resolution $\veps$ for the independent variable. If we take the convolution of $\wig_\rho$ with a Gaussian $G_\veps$ peaked at the origin with small covariance $\veps$, we still get a good qualitative picture. The resulting density $\wig_\rho\ast G_\veps$ will not have exactly the correct marginal distributions, but instead each marginal will appear convolved with a peaked Gaussian as well (see Fig.~\ref{fig:3rand}). The regularized Wigner function $\wig_\rho\ast G_\veps$ is readily computed by multiplying \eqref{FouWig} with a slowly decaying Gaussian $\widehat G_\veps$ before taking the inverse Fourier transform, which then converges reasonably well. All pictures in this paper were generated in this way.
Note that $\wig_\rho=\lim_\veps \wig_\rho\ast G_\veps$ is an alternative definition of $\wig_\rho$.

\begin{figure}[h]
	\includegraphics[width=0.99\columnwidth]{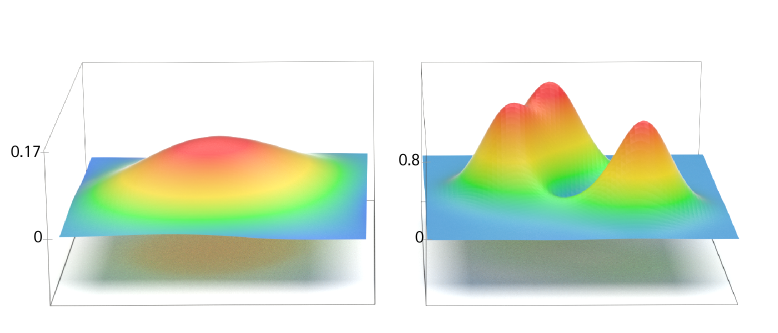}\\[0.3cm]%
	\includegraphics[width=0.99\columnwidth]{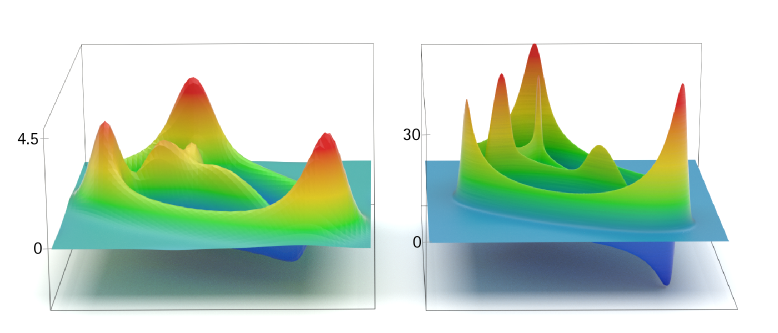}\\[0.3cm]%
	\includegraphics[width=0.99\columnwidth]{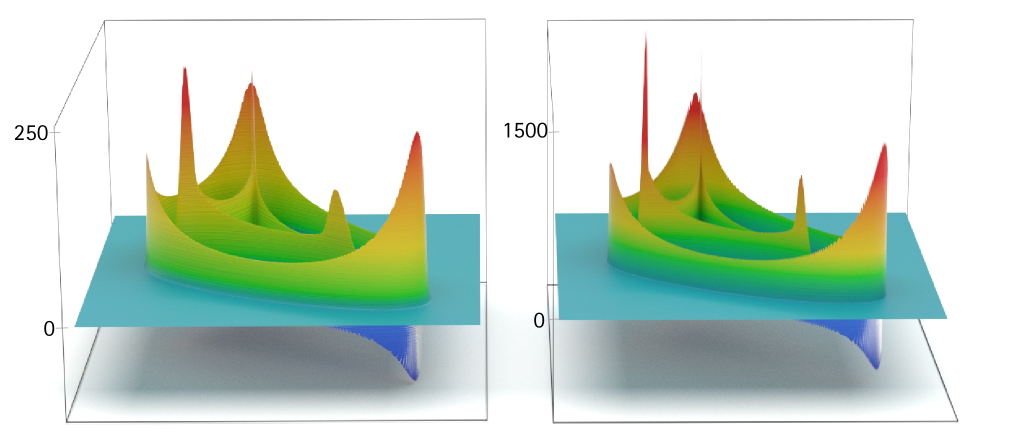}%
	\caption{Dependence on the regularization parameter $\veps$.
		Depicted for $\veps=1,10^{-1},10^{-2},\cdots, 10^{-5}$}
	\label{fig:regularize}
\end{figure}

A distribution like the derivative of the $\delta$-function cannot be plotted. However, visualization is still possible if we take into account that any graphical representation of a function requires a choice of resolution. So at each point we plot instead the average of the distribution over a Gaussian of some width $\veps$, i.e., we take the convolution with such a Gaussian. Equivalently, we multiply $\widehat\wig_\rho(\xi)$ by the Fourier transform of the Gaussian, which is $\exp(-\veps\xi^2)$. The product is then a well-decaying integrable function, and we can take its Fourier transform with standard numerical methods (Fast Fourier transform). This is how all the plots of Wigner functions in this paper were generated. We usually tuned the regularization parameter $\veps$ for best visibility. For too large $\veps$ one gets a featureless bump approximately equal to the regularizing Gaussian. For too small $\veps$ one sees only the singularities in the form of infinitely high walls. Fig.~\ref{fig:regularize} shows the dependence on $\veps$ for the example of two random matrices in $d=4$.

\newpage

\section{Basic properties}
\begin{figure}[ht]
	\includegraphics[width=0.99\columnwidth]{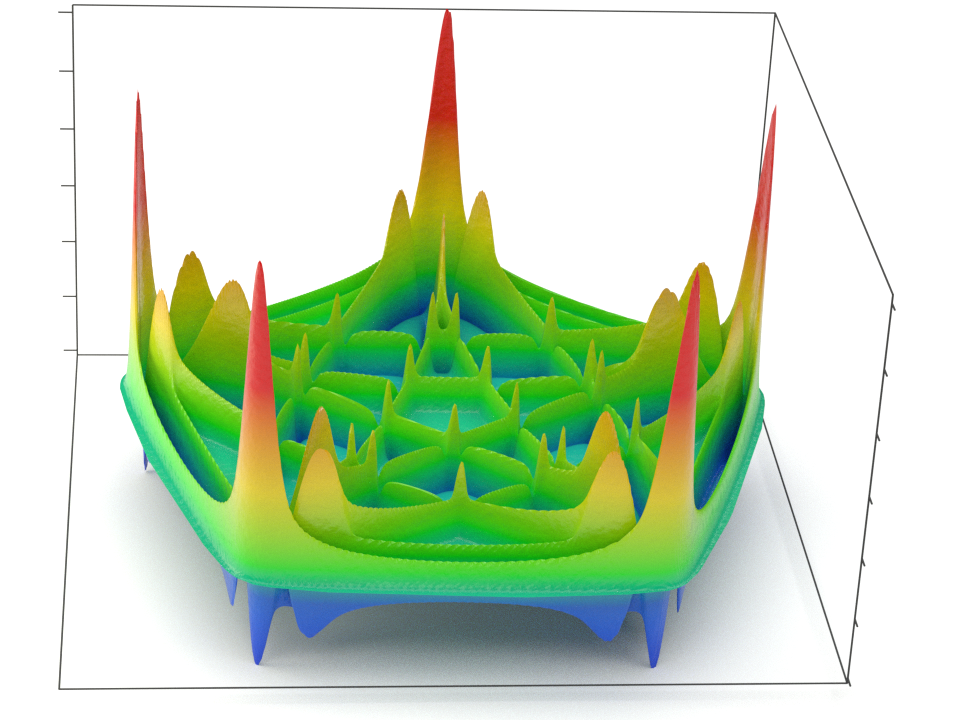}%
	
	\begin{minipage}[c]{0.6\columnwidth}
		\includegraphics[width=0.9\columnwidth]{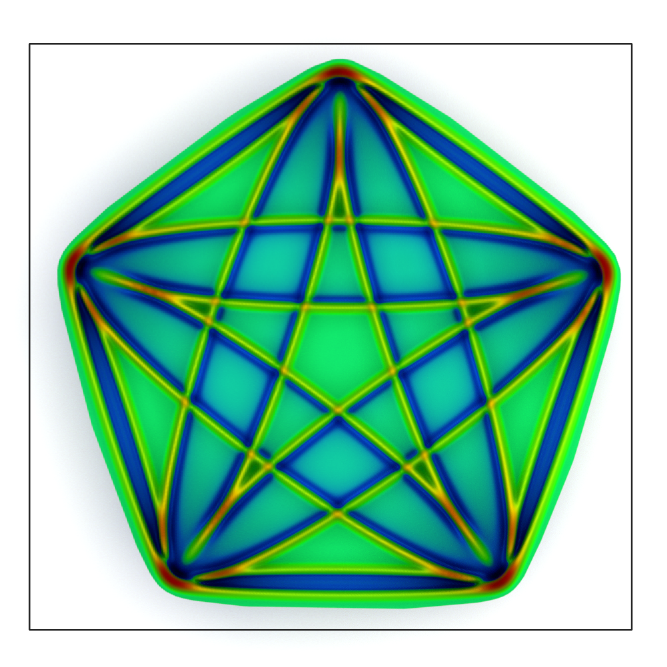}%
	\end{minipage}%
	\begin{minipage}[c]{0.4\columnwidth}
		\caption{\label{fig:penta}\coloronline Wigner function of a pair of operators, which is covariant with respect to the pentagon symmetry.  }
	\end{minipage}
\end{figure}

\begin{figure}[ht]
	\includegraphics[width=0.8\columnwidth]{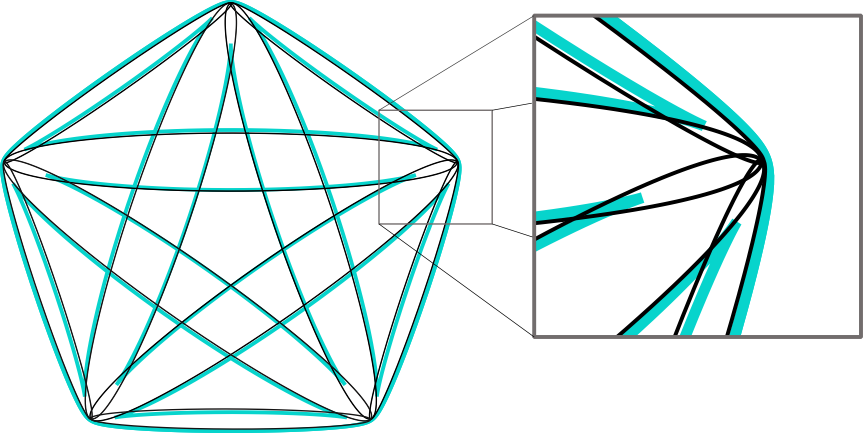}%
	\caption{\label{fig:penta2}\coloronline The same example as Fig.~\ref{fig:penta} as an illustration of the claims for nearly commuting pairs. The connecting ellipses are drawn in thin black and the exact singular manifold $\sing$ in light blue. }\vskip-0.8cm
\end{figure}

We note the following properties in addition to the marginal property \eqref{WigDef}. Proofs are given in the next section.
\begin{itemize}\let\it\bf
\item $\wig_\rho$ is {\it real}, i.e., equal to its complex conjugate distribution. (See \Qref{sec:FourierDef})
	\item $\wig_\rho$ has {\it support} in a compact convex set, the {\it joint numerical range} of the given operators:
	\begin{equation}\label{jnr}
	\jnr=\bigl\{a\in\Rl^n\bigm| a_k=\tr\rho A_k\bigr\},
	\end{equation}
	where $\rho$ now runs over all density operators. That is, for all $\rho$,  $\wig_\rho=0$ outside of $\jnr$. (See \Qref{prop:supp}).
\item The {\it singularities} of $\wig_\rho$ lie on the closure of the set
	\begin{eqnarray}\label{sing}\strut
	\sing&=&\bigl\{a\in\Rl^n\bigm|\bigl.a_k=\brAAket\psi{A_k}\psi;\bigr. \nonumber \\
	&&\strut\hskip40pt\ \norm\psi=1, \xi\cdt A\psi=\lambda\psi \bigr\},
	\end{eqnarray}
where $\lambda$ is a non-degenerate eigenvalue of some $\xi\cdt A$. Outside $\sing$, $\wig_\rho$ is given by an ordinary function, whose value $\wig_\rho(a)$ at some $a\notin\sing$  is given by the $\rho$ expectation of a bounded operator. The convex hull of $\sing$ is $\jnr$. This is described in \Qref{sec:sing}, and the (semi-)algebraic nature of $\sing$ is discussed in \Qref{sec:singAlgebraic}.
\item When the $A_k$ are {\it reducible}, i.e., commute with some hermitian $B\neq\lambda\idty$, $\wig_\rho$ is the sum of the Wigner distributions computed by projecting $\rho$ to the eigenspaces of $B$.
\item When the $A_k$ are finite dimensional matrices, and $\rho$ has full rank, then $\wig_\rho$ is {\it positive} if and only if the $A_k$ commute, in which case $\wig_\rho$ is a sum of $\delta$-functions with $\rho$-dependent weights (see \Qref{sec:pos}).
\item When $n=2$ and $A_1$ and $A_2$ {\it nearly commute}, i.e., are small perturbations of a commuting pair, the $\delta$-peaks of the previous item become approximate, and are connected by singular ellipses. These can be computed to first order by restricting all operators to the two-dimensional eigenspace spanned by the eigenvectors of the commuting pair belonging to the two points. This is illustrated in Fig.~\ref{fig:penta2} (see\Qref{sec:nc}).
\item $\wig$ is {\it covariant} for any symmetry acting on the $A_k$. That is, if $U$ is unitary such that $U^*A_kU=\sum_\ell R_{k\ell}A_\ell$ for some real coefficients $R^g_{k\ell}$, then
	$\wig_{U\rho U^*}(a)=\wig_\rho(Ra)$.
	This is illustrated in Fig.~\ref{fig:penta} (see \Qref{sec:covariant}).
\item For $n\geq3$ the Wigner distribution is generically {\it informationally complete}, i.e., one can reconstruct $\rho$ uniquely from $\wig_\rho$. This also holds when the $A_k$ span the Lie algebra of an irreducible representation (like the angular momentum operators). However, it always fails in finite dimension for $n=2$, because density operators $\rho,\rho'$ with $\rho-\rho'=i\lambda[A_1,A_2^m]$ have the same Wigner distribution.
\end{itemize}

\subsection{Support}
\begin{prop}\label{prop:supp}
$\wig_\rho$ has {\it support} in a compact convex set, the \emph{joint numerical range} $\jnr$ of the given operators:
\begin{equation}\label{jnr}
  \jnr=\bigl\{a\in\Rl^n\bigm| a_k=\tr\rho A_k\bigr\},
\end{equation}
where $\rho$ runs over all density operators.
\end{prop}

The support of a distribution is defined as the smallest closed set $\jnr$ such that for any test function with support disjoint from $\jnr$ the integral vanishes. The definition of joint numerical range used here includes mixed states, so it is automatically compact and convex as the continuous linear image of the state space. In the mathematical literature a definition of joint numerical range based on the pure states only is often used. This is convex for $n=2$ (The Hausdorff-Toeplitz Theorem), and for $n=3$ (except in the qubit case, where the Bloch sphere, in contrast to the full ball, is not convex). Convexity of the pure state joint numerical range fails for $n\geq4$, but its convex hull is always the $\jnr$ defined above.

The core of the proof of  \Qref{prop:supp} 
is the Paley-Wiener-Schwartz Theorem \cite[Thm. 7.3.1]{Hoer}, i.e., the distributional version of the Paley-Wiener Theorem, which links support properties of a function to analyticity properties of the Fourier transform. To get a heuristic idea, consider a function $f$ with support in the half space $\{a|\eta\cdt a\geq c(\eta)\}$, given by the parameters $\eta$ and $c(\eta)$, where $c(\eta)$ is by definition homogeneous ($c(t\eta)=tc(\eta)$ for $t>0$).  Then the function $a\mapsto\exp(-\eta\cdt a)$ is bounded on the support of $f$, so the Fourier integral for $\exp(-\eta\cdt a)f(a)$ converges better than the Fourier integral for $f$, and gives an analytic extension $\widehat f(\xi+i\eta)$ satisfying a bound with a factor $\exp(-c(\eta))$.  The Paley-Wiener-Schwartz Theorem is a precise statement of the converse.

\begin{proof}
In order to control the growth of the analytic extension $\widehat\wig_\rho(\xi+i\eta)$ in the $\eta$-direction uniformly in $\rho$ we will need to control the norm of the operator $\exp(i(\xi+i\eta)\cdt A)$.
Suppose we have the operator inequality
\begin{equation}\label{opineq}
  \eta\cdt A\geq c(\eta)\idty
\end{equation}
for some constant $c(\eta)$, which is equivalent to saying that
\begin{equation}\label{jnrIneq}
  \jnr\subset \{a|\eta\cdt a\geq c(\eta)\}.
\end{equation}
Then $\norm{\exp(-\eta\cdt A)}\leq\exp(-c(\eta))$ and, by the Trotter product formula
\begin{equation}\label{trotter}
  \left\Vert e^{i\xi\cdt A-\eta\cdt A}\right\Vert
  \leq \lim_{n\to\infty}\left\Vert \left(e^{i\xi\cdt A/n}\,e^{-\eta\cdt A/n}\right)^n\right\Vert
  \leq e^{-c(\eta)}.\nonumber
\end{equation}
Hence
\begin{equation}\label{exptype}
  \left|\widehat\wig_\rho(\xi+i\eta)\right|\leq e^{-c(\eta)},
\end{equation}
which means that the support of $\wig_\rho$ is contained in the half space on the right hand side of \eqref{jnrIneq}. Since $\jnr$, as a closed convex set, is the intersection of all half spaces containing it, the support must be contained in $\jnr$. Actually, the last step is already included in the version of the Paley-Wiener-Schwartz Theorem cited above from H\"ormander's book, which is directly based on the function $c(\eta)$, called the 'supporting function' of $\jnr$.
\end{proof}

\subsection{Positivity}\label{sec:pos}
\begin{prop}\label{prop:pos}
When the $A_k$ are finite dimensional matrices, and $\rho$ has full rank, then $\wig_\rho$ is \emph{positive} if and only if the $A_k$ commute,
in which case $\wig_\rho$ is a sum of $\delta$-functions with $\rho$-dependent weights.
\end{prop}

\begin{proof}
Suppose that $\wig_\rho\geq0$. Then it must actually be given by a positive measure, even a probability measure $P$ due to the overall normalization.
Now fix $\xi$ and let $\alpha_i,\alpha_{i+1}$ be two neighbouring eigenvalues of $\xi\cdt A$. Then, by the marginal property, the slice of $\Rl^n$
\begin{equation}\label{strip}
 S_i= \{a| \alpha_i< \xi\cdt a<\alpha_{i+1}\}
\end{equation}
has $P$-measure $0$. In other words, $P$ is supported by the hyperplanes $\xi\cdt a=\alpha_i$. This is true for any direction $\xi$, and taking the intersection over some linearly independent set of directions $\xi$, we find that $P$ is supported on finitely many points. We can then find a direction $\xi$ so that each plane $\xi\cdt a={\rm const}$ contains at most one point. Since $\xi\cdt A$ has at most $d=\dim\HH$ eigenvalues, we find that there are at most $d$ points in the support. Suppose we find this maximal number. Then the corresponding eigenvectors form a basis, and in this basis all $A_k$ are diagonal, so the $A_k$ commute.

Otherwise, we can make this conclusion only for the span of the eigenvectors of $\rho$, and we can draw no conclusion about the commutation of the $A_k$ about the complement of the support of $\rho$. This is why we need the condition that $\rho$ has full rank. The $\rho$-dependent weights are computed as the expectations of $\rho$ in the joint eigenprojections (see previous item).
\end{proof}

It is well-known that Gaussian states do have a positive $(P,Q)$-Wigner functions. By one of the most beautiful results of the theory \cite{Hudson}, these are the only pure states with that property. For mixed states no interesting characterization is known \cite{Broecker}. On averaging over phases space translations (with a sufficiently spread out weight) every state will have positive Wigner function. This is in stark contrast to the above result. However, the proof shows why there is such a great difference: For $A_k$ with discrete spectrum the marginal condition for a positive measure is extremely strong, forcing the measure to vanish on almost all of $\Rl^n$.

\subsection{Singular support}\label{sec:sing}
The singularities of $\wig_\rho$ are its most prominent features. In this section we will locate the {\it singular support}, i.e., the set on whose complement the distribution is a smooth function \cite{reedsimon2}. We will sketch a proof for the finite dimensional case. A rigorous theory for the general case would be highly desirable, especially for infinite dimensional (and possibly unbounded) $A_k$ including absolutely continuous and singular continuous spectrum.

\begin{prop}\label{prop:sing}
For finite dimensional matrices $A_k$, the {\it singularities} of $\wig_\rho$ lie on the closure of the set
\begin{eqnarray}\label{sing}\strut
  \sing&=&\bigl\{a\in\Rl^n\bigm|\bigl.a_k=\brAAket\psi{A_k}\psi;\bigr. \nonumber \\
       &&\strut\hskip40pt\ \norm\psi=1, \xi\cdt A\psi=\lambda\psi \bigr\},
\end{eqnarray}
$\psi$ runs over all eigenvectors of non-degenerate eigenvalue problems for some $\xi\cdt A$.
\end{prop}

To get a heuristic idea, let us first diagonalize each operator $\xi\cdt A$.
\begin{equation}\label{spectral}
  \xi\cdt A=\sum_\mu \alpha_\mu(\xi)\,P_\mu(\xi),
\end{equation}
where $P_\mu(\xi)$, $\mu=1,\ldots,n$, denotes the eigenprojections and $\alpha_\mu(\xi)$ the corresponding eigenvalues.
Then the inverse Fourier integral for $\wig_\rho(a)$ contains, apart from some slowly varying factors the oscillatory phase
\begin{equation}\label{oscillatory}
\exp i(-\xi\cdt a + \alpha_\mu(\xi))
\end{equation}
The stationary phase method for such integrals has it that the dominant contribution to the integral comes from those points at which the gradient of the exponent with respect to the integration variables $\xi$ vanishes, i.e.
$a_k=\partial \alpha_\mu(k)/(\partial\xi_k)$. By first order perturbation theory the right hand side is just $\brAAket\psi{A_k}\psi$ for $\psi$ the corresponding normalized eigenvector, i.e., $a\in\sing$.

\begin{proof}[Sketch of proof for \Qref{prop:sing}]
The diagonalization \eqref{spectral} has to be done only once for each collection of proportional vectors, since, for positive $\lambda$, we have $\alpha_\mu(\lambda\xi)=\lambda\alpha_\mu(\xi)$ and $P_\mu(\lambda\xi)=P_\mu(\xi)$. This suggests to split the variable $\xi=tu$, into a single radius $t\geq0$ and an angular part, i.e., a unit vector $u\in\Rl^n$.

Then the $\veps$-regularized Wigner function is given by
\begin{eqnarray}\label{epsreg}
  \widehat\wig_{\rho,\veps}(tu)&=&\sum_\mu e^{it\alpha_\mu(u)-\veps t^2}\tr(\rho P_\mu(u)) \nonumber\\
  \wig_{\rho,\veps}(a)&=&\sum_\mu \intd u\mskip-12mu\intd tt^{n-1}\,
                 e^{it(\alpha_\mu(u)-a\cdt u)-\veps t^2} \tr(\rho P_\mu(u)) \nonumber\\
      &=& \sum_\mu\intd u h_\veps(\alpha_\mu(u)-a\cdt u)\ \tr(\rho P_\mu(u)),
\end{eqnarray}
$du$ indicates suitably normalized integration with respect to the surface measure of the unit sphere, and $h_\veps$ is the function given by the $t$-integral in the line above.
This is the $n-1$ fold derivative of the corresponding function with $n=1$, which is a Gaussian regularized $\delta$- function. Hence, as $\veps\to0$
the $u$-integration becomes concentrated on the set
\begin{equation}\label{intset}
  \MM(a)=\{u| \alpha_\mu(u)-a\cdt u\},
\end{equation}
which is the intersection of the unit sphere with the manifold with $\alpha_\mu(\xi)=a\cdt\xi$.
Suppose that, for all $u$, $\nabla \alpha_\mu(u)\neq a$. Then by the implicit function theorem $\MM(a)$ is a regular submanifold of the unit sphere. In a neighbourhood of any non-degenerate point $P_\mu(u)$ is an analytic function. In the limit the $n-1$ fold normal derivative of this function is integrated over the submanifold $\MM(a)$, which is a regular integral. Hence the singular support is contained in
\begin{equation}\label{singsupp}
  \bigcup_\mu\{\nabla\alpha_\mu(u)|\,\abs u=1\}.
\end{equation}

[A more careful treatment uses the theory of distributions given by oscillatory integrals \cite[Thm.~IX.47]{reedsimon2}, and the equality of the singular support of such a distribution with its wave front set and in turn with the stationary phase points of the phase function $\xi\mapsto \alpha_\mu(\xi)-\xi\cdt a$.]

By first order perturbation theory, we can evaluate $\nabla\alpha_\mu(\xi)$ as the expectation tuple of the $A_i$ in the normalized eigenvector of $\xi\cdt A$ with eigenvalue $\alpha_\mu$, which is the formula given in \Qref{prop:sing}.
\end{proof}

\subsection{Algebraic nature of the singular support}\label{sec:singAlgebraic}
The singularity set $\sing$ is defined completely in terms of the eigenvalue problems of $\xi\cdt A$, quite independently of the Wigner distribution. It is therefore not surprising that it has been studied in various papers. A set is called {\it algebraic}, if it is the solution set a system of polynomial equations, and {\it semi-algebraic}, if it is the solution set of a system of polynomial equations and inequalities. Then we have:

\begin{prop}\label{prop:semialg}
The singular support $\sing$ is a semi-algebraic set.  For $n=2$ it is even algebraic, but not necessarily for $n\geq3$.
\end{prop}

Clearly, the information about the eigenvalues of the family of operators  $\xi\cdt A$  (sometimes called an ``operator pencil'') is also contained in the algebraic variety $V$ consisting  of the zeros of the homogenous polynomial $\xi_0,\ldots,\xi_n\mapsto \det(\xi_0\idty+\xi\cdt A)$. This is a subset in the space of $\xi$-variables, whereas $\sing$ is a set of tuples of $a$-variables, so lives in the dual vector space. It is therefore natural to consider the {\it dual variety} $V^*$, which is defined in complex algebraic geometry \cite{Harris} as the closure of the set of normal vectors at smooth points.
To see what this gives in the case at hand, let us just look at the points where the rank of $\xi_0\idty+\xi\cdt A$ is reduced by $1$, i.e., $\xi_0\idty+\xi\cdt A$ has a one dimensional kernel. Then there is a unique eigenvector $\psi$ of $\xi\cdt A$ with eigenvalue $-\xi_0$. To get the tangent plane at this point, we check when a curve, say
$(\xi_0(t),\xi(t))=(\xi_0,\xi)+t(\eta_0,\eta)+\order t$ remains in $V$ to first order. By non-degenerate first order perturbation theory, the eigenvalue condition remains satisfied to first order, provided that $-\eta_0=\brAAket\psi{\eta\cdt A}\psi=\eta\cdt a$, where $a$ is the expectation vector of $\psi$. That is, the allowed tangent vectors $(\eta_0,\eta)$ are precisely those orthogonal to $(1,a)$, which defines the normal vector up to a factor. The factor is here chosen to make the first component $1$, which defines the ``affine part'' of the variety $V^*$. So $\sing$ is identified as the affine part of the dual variety.

However, in this description we have implicitly taken all variables to be real. This may be quite different from starting with complex $\xi$, and only restricting to real $a$ at the end. The basic algebraic construction for getting $V^*$ is to eliminate the $\xi$-variables from the equations
\begin{eqnarray}\label{elimDual}
  \det(\xi_0\idty+\xi\cdt A)&=&0 \nonumber\\
  \frac\partial{\partial\xi_0}\det(\xi_0\idty+\xi\cdt A)&=&1, \\
  \frac\partial{\partial\xi_k}\det(\xi_0\idty+\xi\cdt A)&=&a_k, \nonumber
\end{eqnarray}
and take the closure of the resulting set of $a$-tuples. In the real version of this construction, this constitutes an obviously algebraic set of tuples $(\xi_0,\xi,a)$, which is to be projected to just the $a$ variables. This operation takes semi-algebraic sets to semi-algebraic sets, as does the closure, which shows that indeed $\sing$ is semi-algebraic.
On the other hand, consider a purely algebraic elimination process, which will be the same in the complex and in the real case, and hence implicitly includes cases of complex $\xi$ in \eqref{elimDual}. This leads to the dual variety $V^*$, which is again an algebraic variety. Its algebraic definition as the zero set of a polynomial automatically describes a closed set, but its intersection with the reals may now contain additional points. In such a case a purely algebraic description of $\sing$ fails, and we need some inequalities in addition to polynomial equations to characterize it.

\begin{figure}[h]
	\includegraphics[width=0.99\columnwidth]{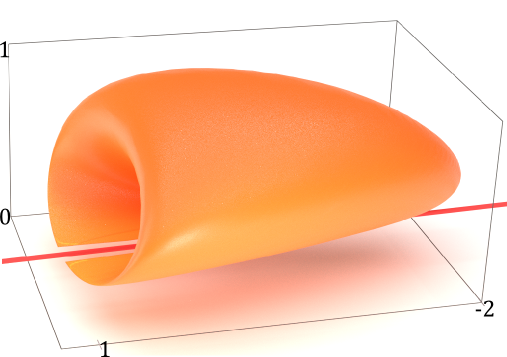}%
	\caption{Zero set of the polynomial \eqref{gpoly}, the union of $\sing$ (shell-like body) and the red straight line. Parametrization of the body is by
             \eqref{areplace}, with $y$ artificially restricted to leave some space around the straight line for visibility.}
	\label{fig:semialg}
\end{figure}

\begin{figure}[h]
	\includegraphics[width=0.99\columnwidth]{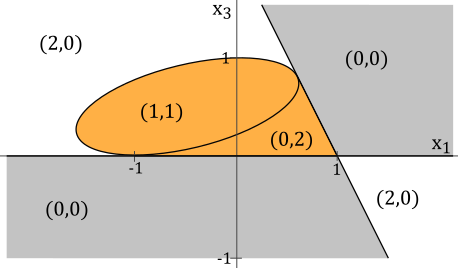}%
	\caption{Sketch of proof that \Qref{fig:semialg} is a complete representation of the zero set. Setting $g(a_1,a_2,a_3)=h(a_2^2; a_1,a_3)$, the diagram shows the regions in the $a_1$-$a_3$ parameter plane (labelled with $(n_-,n_+)$), where the quadratic polynomial $u\mapsto h(u;a_1,a_3)$ has $n_-$ negative and $n_+$ positive zeros. The yellow area, where there is at least one positive root is a side view of \Qref{fig:semialg}.}
	\label{fig:semialgline}
\end{figure}

It was shown by Kippenhahn \cite{Kippen} that for $n=2$ this phenomenon does not happen, so that the algebraic construction correctly describes $\sing$. A more modern and more detailed version is provided by \cite{Chien}, see also \cite{Weis}. In \cite{Chien} there is also a counterexample for $n=3$. We include it here, pointing out more explicitly how complex $\xi$ in \eqref{elimDual} enter to get points outside $\sing$. The operators are
\begin{equation}\label{mats}
A_1={\scriptstyle\left(
\begin{array}{ccc}
 1 & 0 & 0 \\
 0 & {-1} & 1 \\
 0 & 1 & 0 \\
\end{array}
\right)},\
A_2={\scriptstyle\left(
\begin{array}{ccc}
 0 & 0 & 1 \\
 0 & 0 & 0 \\
 1 & 0 & 0 \\
\end{array}
\right)},\
A_3={\scriptstyle\left(
\begin{array}{ccc}
 0 & 0 & 0 \\
 0 & 0 & 0 \\
 0 & 0 & 1 \\
\end{array}
\right)}\nonumber
\end{equation}
Then the eigenvalues $-\xi_0$ are the zeros of
\begin{eqnarray}\label{det}
  f(\xi_0,\xi)\hskip-20pt&&\hskip20pt=\det(\xi_0\idty +\xi\cdt A) \\
  &&=\xi_{0}^3+\xi_{0}^2 \xi_{3}-2 \xi_{0} \xi_{1}^2-\xi_{0} \xi_{2}^2-\xi_{1}^3-\xi_{1}^2 \xi_{3}+\xi_{1} \xi_{2}^2,\ \nonumber
\end{eqnarray}
The algebraic elimination of the $\xi$-variables from \eqref{elimDual} leads to the irreducible polynomial
\begin{eqnarray}\label{gpoly}
  g(a)&=&4 a_{3}^2 \left(a_{1}^2-2 a_{1} a_{3}+5 a_{3}^2  +2 a_{1}-6 a_{3}+1\right)+ \nonumber\\
           &&+4 a_{3} (2 a_{3}-a_{1}-1)a_2^2+ a_{2}^4.
\end{eqnarray}
The affine part of the dual complex variety is thus described by $g(a)=0$. This is the best purely algebraic description of $\sing$ that we can get. Indeed from any small piece of the surface we can reconstruct this polynomial. The resulting surface $\sing$ is shown in \Qref{fig:semialg}.
But it has a new feature namely the infinite line $(a_1,0,0)$ with arbitrary $a_1\in\Rl$. Since this is an unbounded line, it clearly does not belong to the closure of $\sing$. One can verify that it is the only addition to $\sing$ in the zero set of $g$ (see \Qref{fig:semialgline} for a sketch).

How do we get that strange line now? The only points on it, which can arise as directly as solutions of the system \eqref{elimDual} are $a_0=\pm1$. The other points must arise from a closure operation. Consider the parameterized set of points on $V$
\begin{equation}\label{xireplace}
  {\textstyle\begin{pmatrix}\xi_0\\\xi_1\\\xi_2\\\xi_3\end{pmatrix}}
     ={\textstyle\begin{pmatrix}y -1\\y \\x  (2 y -1)\\  x^2\left(2 y -1\right)+1-y^2 \end{pmatrix}},
\end{equation}
corresponding via \eqref{elimDual} to
\begin{eqnarray}\label{areplace}
  {\textstyle\begin{pmatrix}a_1\\a_2\\a_3\end{pmatrix}}
     &=&\frac{1}{1+x^2+y^2}{\begin{pmatrix}{x^2-y^2+2y}\\{-2x}\\1 \end{pmatrix}} \\
     &&\longrightarrow\frac{1}{x^2+y^2}{\textstyle\begin{pmatrix}{x^2-y^2}\\{0}\\0 \end{pmatrix}},
\end{eqnarray}
where the arrow indicates the scaling limit $x\mapsto\lambda x$, $y\mapsto\lambda y$, and $\lambda\to\infty$. When $x,y\in\Rl$ this shows that $a=(a_1,0,0)\in\sing$, when $a_1\in[-1,1]$. However, if we allow complex parameters, say $y\mapsto i\lambda y$, we get in addition the inverses of the same $a_1$, i.e., the rest of the real line.
Hence the purely algebraic description of $\sing$ fails, and inequalities are needed in addition.

Another characterization of $\sing$ is as the set of expectation tuples $\brAAket\psi{A}\psi$ for which the map $\psi\mapsto \brAAket\psi{A}\psi\in\Rl^n$ has reduced rank $\leq(n-1)$. For $n=2$ this investigated in \cite{Joswig}.

\subsection{Lack of $L^2$-isometry}
One well-known property of the standard Wigner function is that it is unitary from the Hilbert Schmidt class to the square integrable functions on phase space:
\begin{equation}\label{L2iso}
  \int \frac{dp\,dq}{2\pi}\ \overline{\wig_\rho(p,q)}\, \wig_\sigma(p,q)=\tr(\rho^*\sigma).
\end{equation}
for arbitrary trace class operators $\rho$ and $\sigma$. Equivalently, since the inverse Fourier transform on $L^2$ is unitary, this formula  holds for $\widehat\wig$ instead of $\wig$. Such a relation does not hold in the general case, and in this section we will consider the obstructions to such a formula for any tuple of finite dimensional matrices.

The key property for \eqref{L2iso} is the so-called square integrability of the Weyl representation. Phrasing this condition in the language used here, this property means that, for any pair of vectors $\phi,\psi$, the matrix element $\brAAket\phi{\exp(i\xi\cdt A)}\phi$ is square integrable as a function of $\xi$. It is shown in \cite{QHA} how this allows one to consider the trace class as a convolution algebra for which the ``Weyl transform'' $\rho\mapsto\widehat\wig_\rho$ acts as the ``Fourier'' transform turning it into an algebra of functions. The Wigner function is then the concatenation of the Weyl transform and the ordinary Fourier transform, making unitarity obvious, and also the property that the convolution of two Wigner functions is a positive integrable function.

But in finite dimension the matrix element $\brAAket\phi{\exp(i\xi\cdt A)}\phi$ and its square are almost periodic functions in every direction $\xi$. Hence every contribution to the integral is (almost) repeated infinitely often, and the integral cannot converge. More generally, discrete spectrum prevents this integrability (at least for $\phi,\psi$ associated with eigenvectors of some $\xi\cdt A$). So the inverse Fourier transform \eqref{FouFouWig} is not defined pointwise, and there is usually no {\it operator} $\Delta(a)$ such that $\wig_\rho(a)=\tr(\rho \Delta(a))$. Note that this fits well the description of the singular support: Precisely when $a\in\sing$ we have periodic behaviour of the Fourier integral for some vectors and some direction $\xi$.

To contrast this again with the standard Wigner function: In that case $\Delta(a)$ is (up to a factor $2^n$) the operator of phase space inversion at the point $a$ \cite{Grossmann}, and hence all these operators are bounded by $2^n$. By \eqref{L2iso} they formally satisfy the normalisation $\tr(\Delta(a)\Delta(a'))=\delta(a-a')$, but again there is no chance of this (or \eqref{L2iso}) to make sense, since all $\wig_\rho(A)$ (and hence $\Delta(a)$) have their distributional singularities on the same set $\sing$.

One might try to work with just those regions (away from $\sing$) where the general Wigner functions are regular and, possibly, bounded operators $\Delta(a)$ exist. However, these operators typically diverge when $a$ approaches $\sing$, and there is no way to split $\wig_\rho$ into a regular part plus a purely singular distribution. An example of this is discussed in Sect.~\ref{sec:qbitn=2}.

One might also try to generalize \eqref{L2iso} to set up some bilinear expression, or perhaps a test function $\tau$ of two arguments such that, formally:
\begin{equation}\label{bilitr}
  \int da\, da'\ \tau(a,a')\ \wig_\rho(a)\wig_\sigma(a')= \tr(\rho\sigma)
\end{equation}
But alas, this will also not work in general: It would imply that $\rho$, which can clearly be reconstructed from all its scalar products $\tr(\rho\sigma)$ with varying $\sigma$, could be reconstructed from $\wig_\rho$ (informational completeness). But as discussed in Sect.~\ref{sec:infocomp} this property fails in general.

Of course, it would be interesting to find operator tuples $A$ (possibly unbounded, possibly with continuous spectrum) for which such formulas are possible.

\subsection{Reducibility}\label{sec:reduce}
We call a tuple $A_1,\ldots,A_n$ {\it irreducible}, if the only linear operators commuting with all $A_k$ are the multiples of $\idty$. This is equivalent to saying that the weakly closed algebra of operators generated by the $A_k$ is the algebra $\BB(\HH)$ of all bounded operators. In the finite dimensional case one can omit the closure here. In the reducible case we have a joint diagonal block decomposition of all $A_k$, and the Wigner function is insensitive to the off-diagonal matrix elements of $\rho$ with respect to that decomposition:

\begin{prop}\label{prop:red}
Suppose all $A_k$ commute with the a family of orthogonal projections $Q_\ell$ with $\sum_\ell Q_\ell=\idty$.
Then
\begin{eqnarray}
 \widehat\wig_\rho(\xi) =\sum_\ell \widehat\wig_{Q_\ell\rho Q_\ell}(\xi).  \nonumber
\end{eqnarray}
\end{prop}

Note that in this case the support of $\wig_\rho$ is contained in the union of the corresponding joint numerical ranges $\jnr_\ell$, which may be much smaller than $\jnr$, their convex hull. In other words, \Qref{prop:supp} only gives an upper bound on the support.

Reducibility in the above sense also implies that the determinant polynomial $\det(\xi_0\idty+\xi\cdt A)$ factorizes, i.e., the variety $V$ is ``reducible''. Kippenhahn conjectured this to be the only case. However, that is only true in small ($d\leq5$) matrix dimension. There is a counterexample \cite{Laffey} in $d=8$ dimensions with $n=2$.

Rather than commutation with a linear operator we can consider an antilinear operator. This also leads to a lack of informational completeness, i.e., $\rho$ cannot be reconstructed from $\wig_\rho$. In order to avoid trivial reductions of a kind we already know the following Proposition assumes irreducibility.

\begin{prop}\label{prop:infoNoReal} Let $A_1,\ldots,A_n$ be an irreducible set of hermitian $d\times d$-matrices, all commuting with some non-zero antilinear operator $\Theta$.
Then all Weyl-ordered moments also lie in $\MM_\Theta=\{X|\,X^*{=}X,\ \Theta X{=}X\Theta\}$. Moreover, $\Theta^2=\lambda\idty$, and there are two possibilities:
\begin{itemize}
  \item[(1)] $\lambda>0$, and $\dim_\Rl(\MM_\Theta)= d(d+1)/2$, and
  \item[(2)] $\lambda<0$, and $\dim_\Rl(\MM_\Theta)= d(d-1)/2$
\end{itemize}
\end{prop}

\begin{proof}
The statement that $\MM_W\subset\MM_\Theta$ is trivial.

The linear operator $\Theta^*\Theta$ also commutes with all $A_k$, hence must be a multiple of the identity. We have excluded $\Theta=0$, so after multiplication by a positive normalization factor $\Theta$ becomes unitary.
Since $\Theta^2$ also commutes with the $A_k$ it must be $\lambda\idty$, with $\abs\lambda=1$ by antiunitarity, and $\lambda\in\Rl$ because $(\lambda\idty)\Theta=\Theta^3=\Theta(\lambda\idty)$. Hence $\lambda=\pm1$.

When $\Theta^2=+\idty$, we can choose an orthonormal basis of the real Hilbert space $\{\phi|\Theta\phi=\phi\}$, which will then also be a basis for $\HH=\Cx^d$. In this basis $\Theta$ is just complex conjugation, and $X\in\MM_\Theta$ is equivalent to $X$ being a real and symmetric matrix in that basis. This is uniquely specified by $d$ real diagonal elements, and $d(d-1)/2$ off-diagonal matrix elements, altogether $d(d+1)/2$ parameters.

When $\Theta^2=-\idty$, each $\psi$ mapped to an orthogonal vector ($\psi{\perp}\Theta\psi$), because $\braket\psi{\Theta\psi}=\braket{\Theta^2\psi}{\Theta\psi}=-\braket\psi{\Theta\psi}$. This implies that $d$ is even, and we can build a basis by including with each vector $\psi_j$ the vector $\psi_{d/2+j}:=\Theta\psi_j$. With respect to such a basis we can block decompose $\Theta$ and an arbitrary element $X\in\MM_\Theta$ as
\begin{equation}\label{XTheta}
  \Theta=\begin{pmatrix}0&K\\-K&0\end{pmatrix} \qquad X= \begin{pmatrix}A&B\\C&D\end{pmatrix},
\end{equation}
where $K$ stands for complex conjugation. Then $X\in\MM_\Theta$ translates to $A=A^*=KDK$, and $C=B^*=-KBK$. Thus we have to choose an arbitrary hermitian $(d/2)\times(d/2)$-matrix $A$ (requiring $(d/2)^2$ real parameters), and an antisymmetric one, $B$, with $2(d/2)(d/2-1)/2$ real parameters, together $d^2/2-d/2$.
\end{proof}

Thus in either case, the number of parameters is much smaller than $d^2$, which would be required for informational completeness (see \Qref{sec:infocomp}).

\subsection{Strict convexity}
\begin{prop}\label{prop:strconv}
For $n\leq3$ the boundary of $\jnr$ is generically strictly convex and belongs to $\sing$.
\end{prop}

Thus the typical appearance of a Wigner function for two operators is that of a walled-in convex city rising above a plain. The city walls come from the singularity, and can be made arbitrarily high by choosing a a small regularization parameter $\veps$. Together with the walls there is always a moat. More formally, the singularity cannot be positive only, because that would clash with the marginal property. The moat is on the inside of the wall. It is not entirely clear how to define this as a property of a distribution. But this property is exhibited by all our graphs. For a heuristic argument consider a supporting hyperplane
\begin{equation}\label{supplane}
 \bigl\{a\bigm|\xi\cdt a=m(\xi)\bigr\},
\end{equation}
where $m(\xi)=\max\{\xi\cdt x|x\in\jnr\}$. Then the integral of the regularized Wigner function over the half space $\xi\cdt a\geq0$ can be computed from the marginal property, and will be positive if the regularizing function is positive.

The crucial property implying strict convexity is that for every $\xi$ the maximal eigenvalue of $\xi\cdt A$ is non-degenerate.
Then the supporting hyperplane \eqref{supplane} meets $\jnr$ in a unique point and, moreover, this point is the expectation tuple belonging to a unique pure state. Thus, according to \eqref{sing}, it belongs to $\sing$.

\begin{figure}[t]
\includegraphics[width=0.6\columnwidth]{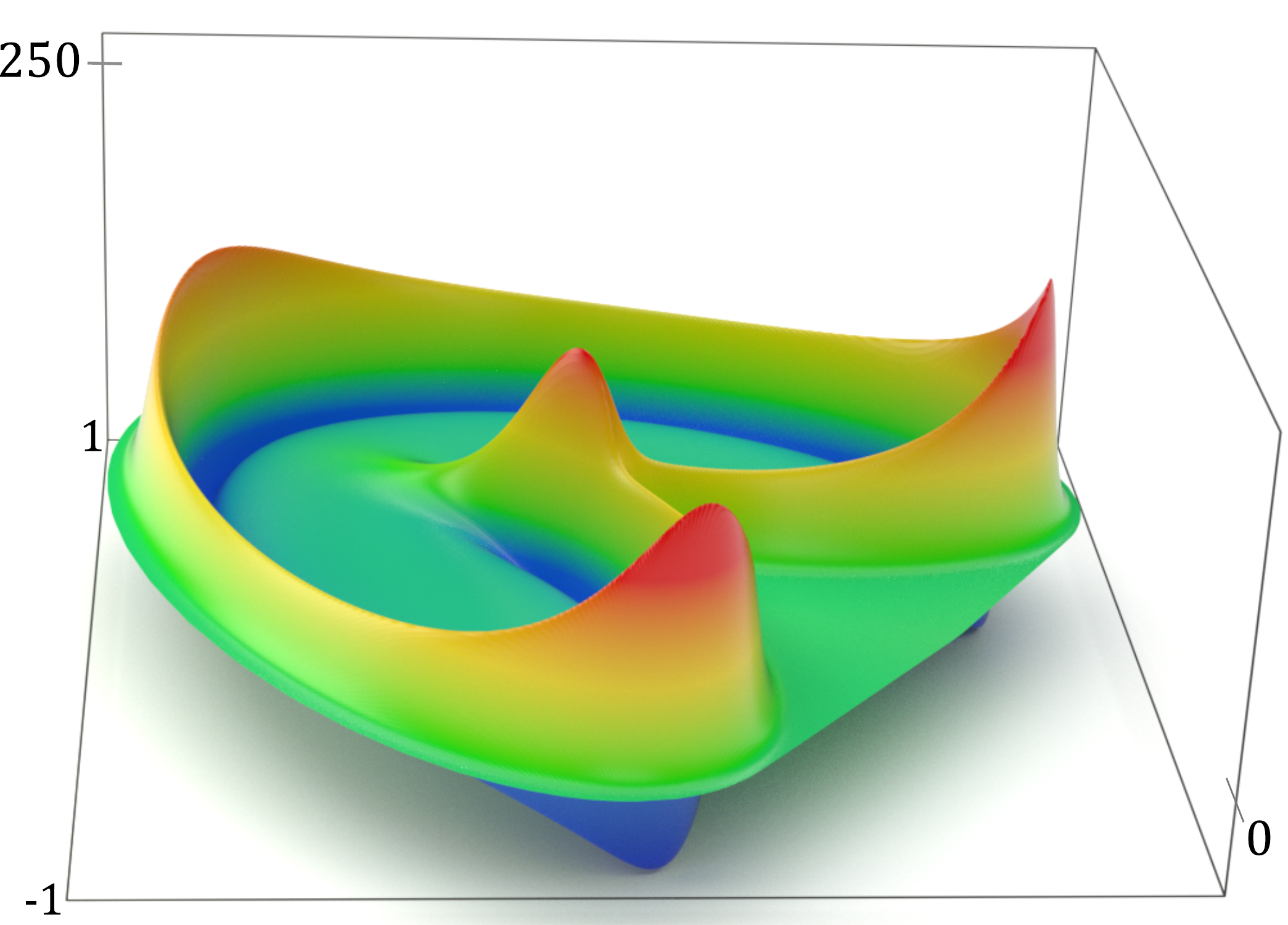}
\includegraphics[width=0.35\columnwidth]{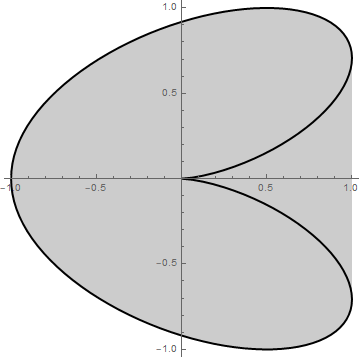}
\caption{\coloronline Left: Wigner function of two $3\times3$-matrices \eqref{heartMatrix}.
    Right: The singularity curve $\sing$, the quartic \eqref{heartquart}. Grey: the range $\jnr$. }
\label{fig:heart}
\end{figure}

Before showing this feature to be generic, let us give some examples how it may fail. The simplest are reducible tuples, where the straight lines arising from the convex hull of widely separated $\jnr_\ell$ clearly disprove strict convexity. However, reducibility is non-generic, because it is typically destroyed by an arbitrarily small perturbation. A less trivial counterexample (depicted in \Qref{fig:heart}) is given by
\begin{equation}\label{heartMatrix}
  A_1=\begin{pmatrix} 0&0&1\\0&1&0\\1&0&0\end{pmatrix}, \qquad A_2=\begin{pmatrix} 0&0&0\\0&0&1\\0&1&0\end{pmatrix},
\end{equation}
and is taken from \cite[Sect.~4.1]{Henrion}. Its characteristic feature is the double eigenvalue $1$ of $A_1$. The singularity curve is the quartic
\begin{equation}\label{heartquart}
  4 a_1^3 + 4 a_1^4 - 27 a_2^2 - 18 a_1 a_2^2 + 13 a_1^2 a_2^2 + 32 a_2^4=0.
\end{equation}

We now have to establish that the non-degeneracy of maximal eigenvalues is generic. We go for the even sharper statement that {\it all} eigenvalues of all $\xi\cdt A$ are simple. This was discussed by von Neumann and Wigner \cite{vNWig} in 1929. They counted the dimension of the manifold of hermitian $d\times d$-matrices without degenerate eigenvalues, and found it to be $d-3$.  Hence for $n\leq3$, the $\leq2$-dimensional families of operators $\xi\cdt A$ (we can take $\abs\xi=1$), will have no degenerate eigenvalues. For tuples with larger $n$ one expects to see a failure of strict convexity, and this is discussed in detail in \cite{Schimanski}.

\subsection{Nearly commuting pairs}\label{sec:nc}
\begin{figure}[h]
	\includegraphics[width=0.99\columnwidth]{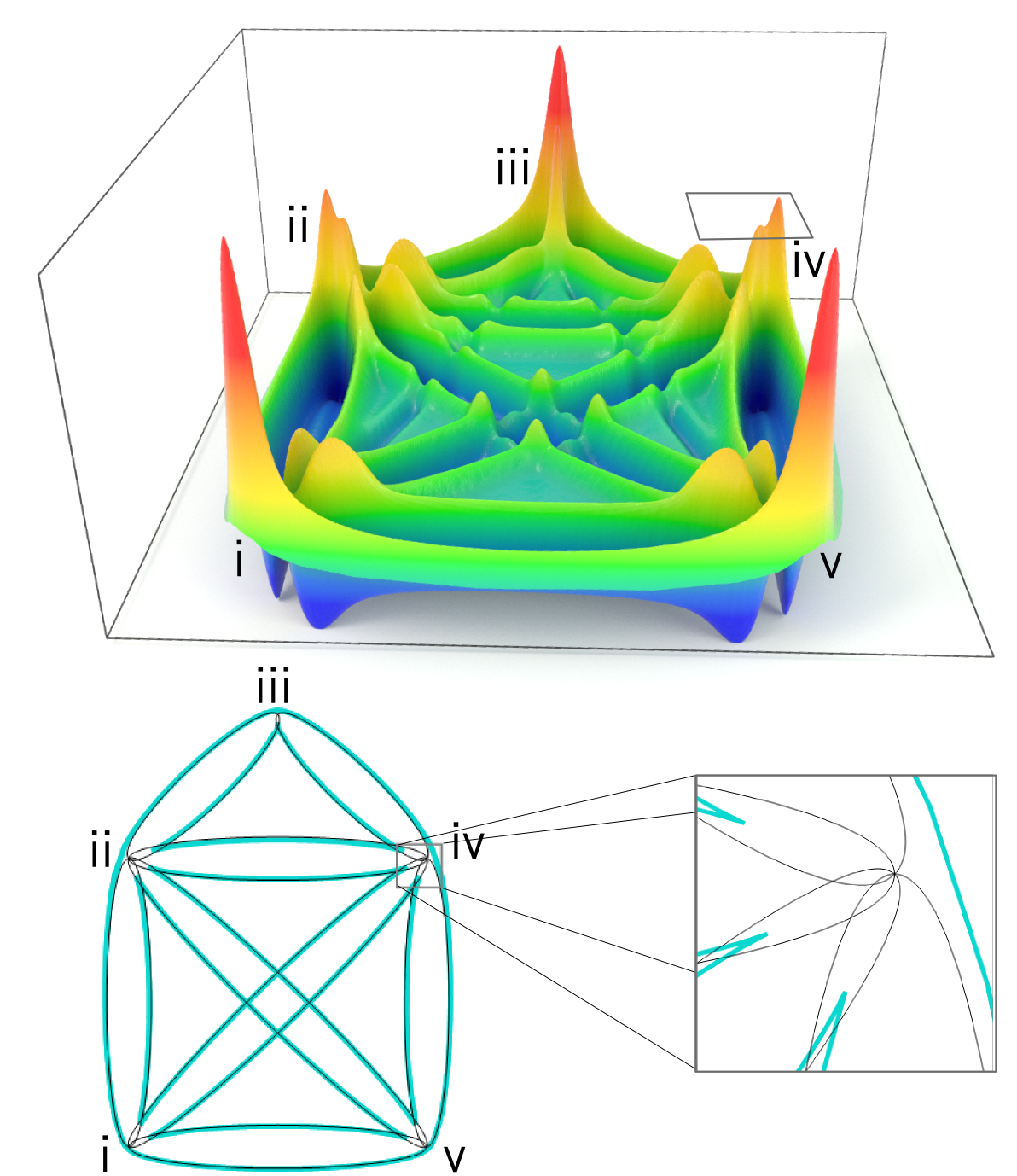}\\
	\def\svgwidth{0.5\textwidth}
	\caption{\coloronline Wigner function (top) and singular set (bottom) of a pair of matrices illustrating \Qref{prop:nc}.
    The singular set $\sing$ is represented in thick light blue. Thin lines are the ellipses described in \Qref{prop:nc}.
    Bottom right: blow-up of the upper right corner indicated in the left diagram. }
	\label{fig:Niko}
\end{figure}

For commuting pairs the Wigner function is a collection of $\delta$-peaks. Now consider a small perturbation, so that $A_1$ and $A_2$ {\it nearly commute}. Then the $\delta$-peaks become approximate, and one observes numerically that the former peaks become connected by singular ellipses. An example of this is shown in Fig.\ref{fig:Niko}.
For simplicity, we choose a basis in which the unperturbed parts are jointly diagonalized. So that the perturbation consists in the off-diagonal elements. Then we claim the following

\begin{prop}\label{prop:nc}
Let $A_1,A_2$ be hermitian matrices, so that the off-diagonal elements are small, say of order $\veps$.
Then singular set $\sing$ is close to the union of ellipses, which are the singular sets of corresponding $2\times2$-submatrices.
\end{prop}

With a small perturbation of a commuting operator tuple one might expect the eigenvectors to be close in first order to those of the commuting system, and hence also the expectation tuples of the perturbed eigenvectors. However, the singular manifold $\sing$ of such a perturbation is not at all concentrated near the unperturbed $\delta$-points that we know from the previous paragraph. It turns out that the explanation is in terms of the well-known phenomenon of {\it avoided level crossing}.

\begin{figure}[h]
	\includegraphics[width=0.99\columnwidth]{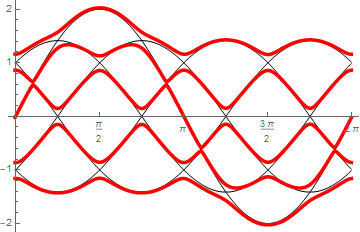}%
	\caption{Eigenvalue curve of the operator family $C(t)=A_1(\veps)\cos t+A_2(\veps)\,\sin t$ as a function of $t$,
     for a commuting pair ($\veps=0$, thin line), and the nearly commuting pair ($\veps>0$, thick red) considered in Fig.~\ref{fig:Niko}}
	\label{fig:avoidX}
\end{figure}

To illustrate this effect, we begin with two commuting operators $A_1,A_2$, the one-parameter family of operators $C(t)=A_1\cos t+A_2\,\sin t$.
The eigenfunctions (say, $\ket\mu$) are the same for all $t$, and the eigenfunctions $c_\mu(t)=a_{1,\mu}\cos t+a_{2,\mu}\sin t$ simple sinusoidal functions (see the thin lines in Fig.~\ref{fig:avoidX}). These curves contain all the information about $\sing$, also in general. Indeed, for any $t$ we can consider a normalized eigenvector $\psi$ of $C(t)$ at some degenerate point. Then by the eigenvalue property the expectation tuple $(a_1,a_2)$ satisfies $c_\mu(t)=a_{1}\cos t+a_{2}\sin t$. On the other hand, by the perturbation theory of the non-degenerate eigenvalue we get the $t$-derivative
$\dot c_\mu(t)=-a_1\sin t+a_2\cos t$. Hence,
\begin{equation}\label{afromccdot}
  \begin{pmatrix}a_1\\a_2\end{pmatrix}= \begin{pmatrix}\cos t&-\sin t\\\sin t&\cos t\end{pmatrix}\begin{pmatrix}c_\mu(t)\\\dot c_\mu(t)\end{pmatrix}.
\end{equation}
The only thing special about the commuting case is that the $a_i$ do not depend on $t$.

We now turn on a small perturbation so that $A_i(\veps)=A_i+\veps A_i'$ with $\veps$ small. The resulting eigenvalue curves of $C(t,\veps)$ are shown as thick lines in Fig.~\ref{fig:avoidX}, and exhibit various avoided crossings. Between the crossings the perturbed eigenvalue curves (and their derivatives) are close to the unperturbed ones, so the corresponding expectation pairs remain close to one of the unperturbed points $\vec a_\mu$. In the vicinity of the crossings however, one sees a switch between curves and hence between two different $\vec a_\mu$. This is possible, because, while the eigenvalues stay close to the unperturbed ones, their $t$-derivatives don't. This resolves the apparent paradox mentioned in the opening paragraph of this section.

For a more quantitative theory, it is helpful to consider the avoided crossing phenomenon as a case of degenerate perturbation theory in two parameters.
Imagine a family of curves parametrized by the perturbation $\veps$ and drawn at height $\veps$ above the paper plane. Let us focus on a crossing that happens at $t=0$ (For other points we may just rotate the original operators and the parameter plane). This is to say, there are two indices $\mu,\nu$ for which $a_{1,\mu}=c_\mu(0)=c_\nu(0)=a_{1,\nu}$, i.e., $A_1$ has a degenerate eigenvalue. We look at the first order perturbation theory of this eigenvalue
with respect to the two parameters $t$ and $\veps$, i.e., in this order,
\begin{equation}\label{2parpert}
  C(t,\veps)=A_1+\veps A_1'+t A_2.
\end{equation}
Although one of the standard textbooks on perturbation theory \cite[Sect.~II.7]{Kato} turns away in disgust from multi-parameter perturbations (see, however, \cite{Baumgaertel}), we can extract what we need by simple application of the standard one-parameter theory. Along any straight line in $(\veps,t)$-parameter space, for example by fixing $t/\veps$ and taking $\veps$ as the üperturbation parameter, we can apply the standard theory. We conclude that along this line the degenerate eigenvalue of $A_1$ branches, and the slope of the two branches are the eigenvalues of $P(\veps A_1'+(t/\veps)A_2)P$, where $P$ is the projection onto the degenerate eigenspace, spanned by $\ket\mu$ and $\ket\nu$.
The crucial fact is now that the perturbation directions $A_1'$ and $A_2$ usually do not commute. Therefore the slope of the branching is not linear in the two perturbation parameters. However, since that property holds along any single line we get (to first order) a conical surface.

Since the projection $P$ is two-dimensional, we can say more: We can write these operators as $PA_1'P=x_0\idty+\vec x\cdt\vec\sigma$, where $x_0,x_1,x_2,x_3$ are real and the scalar product is between $\vec x=(x_1,x_2,x_3)$ and the vector $\vec\sigma$ of Pauli matrices.
Writing $PA_2P$ similarly with coefficients $y_i$ we find the eigenvalues of \eqref{2parpert} to be
\begin{equation}\label{2parpev}
  c(t,\veps)=a_{1,\mu}+ (\veps x_0+ty_0)\pm\abs{\veps\vec x+t\vec y}.
\end{equation}
As a function of $(\veps,t)$ this is an elliptic double cone, in the simples case ($A_1'$ and $A_2$ distinct Pauli matrices) a section of the light cone of SRT.
Coming back to the family of curves drawn with the perturbation parameter $\veps$ as the vertical direction: Near a crossing we expect to see the intersection of a double cone with the $\veps$-plane, which is a pair of hyperbolas.

Drawing the expectation curves determined by the thick lines in Fig.~\ref{fig:avoidX} then gives ellipses. But we do not have to through the detailed computation, because what we need to compute is exactly the same as for the singular manifold of the Wigner function of the pair $(PA_1P,PA_2P)$. These are obviously ellipses, because $\jnr$ is just a projection of the Bloch ball, and $\sing$ is its boundary.
This is the description given in the text.

We close this section with a remark  on the specific choices made for the diagrams. The basic shape of the five points and the connecting lines (ellipses in the limit $\veps\to0$ is familiar to German children, since it comes with a rhyme and the challenge to draw it in one go without lifting the pencil. This fixes the commuting $A_1,A_2$. The perturbations were chosen off-diagonally to connect appropriate corners. Strengths of these matrix elements were chosen deterministically, and  so that after rotation one would get the same degenerate perturbation problem, i.e., ellipses of roughly the same width. Of course, the matrix element connecting the top vertex to the bottom ones were chosen to be zero. Since the perturbation chosen is already fairly large, however, this gives second order effects, and one sees thin ellipses appearing around these ``unwanted'' diagonals. In other words, the corresponding crossings in Fig.~\ref{fig:avoidX} are also ``avoided'' even if that is not apparent. To get smoother pictures we therefore omitted some points around these crossings, and replacing them by straight lines.

\subsection{Informational completeness}\label{sec:infocomp}
By informational completeness we mean the property that the Wigner function determines the state, i.e., that the map $\rho\mapsto\wig_\rho$ is injective.
This is a key feature of the standard Wigner function. However, in the finite case we have:

\begin{prop}\label{prop:infcomp}
For every pair of hermitian matrices $(A_1,A_2)$ informational completeness fails.
For generic triples $(A_1,A_2,A_3)$ it holds.
\end{prop}

It is convenient to rephrase informational completeness in terms of Weyl-ordered moments: $\rho$ is determined by $\wig_\rho$ if and only if the Weyl-ordered moments are weakly dense in $\BB(\HH)$, or in finite dimension: when the moments span the full matrix algebra. Let us denote by $\MM_W$ the linear space of operators obtained in this way. Then informational completeness is equivalent to $\MM_W=\MM_d$.

To establish the negative statement in \Qref{prop:infcomp} it suffices to point out some hermitian element that has zero trace with all Weyl-ordered moments of two operators. This will be $i[A_1,A_2]$.  This is easily seen as follows. For arbitrary $R\in\Nl$, and $\lambda\in\Rl$:
\begin{eqnarray}\label{comutator}
  \tr\bigl( i[A_1,A_2]&&(A_1+\lambda A_2)^R\bigr) =  \nonumber \\
  &&= i\tr\bigl([A_1+\lambda A_2,A_2](A_1+\lambda A_2)^R\bigr)\nonumber \\
  &&= i\tr\bigl(\bigl[(A_1+\lambda A_2)^{R+1},A_2\bigr]=0.
\end{eqnarray}
So if $[A_1,A_2]\neq0$, informational completeness fails. When $[A_1,A_2]=0$, it fails anyhow, because then only diagonal elements of $\rho$ in the common eigenbasis of $A_1$ and $A_2$ can be distinguished. The reason that this failure does not extend to the standard case is that two density operators can never differ by $i[P,Q]=\idty$.

There is a second straightforward argument for \Qref{prop:infcomp}, which shows that there must be more than just the commutator. When $A_1$ and $A_2$ are both real in the same basis, the same dimension estimate follows from \Qref{prop:infoNoReal}. However, this is just a special case, since usually no such basis can be found.

\begin{lem}\label{lem:cayleyH}
For every pair of hermitian $d\times d$-matrices $(A_1,A_2)$ the real vector space spanned by the Weyl-ordered moments satisfies
\begin{equation}\label{dimMMW}
    \dim_\Rl \MM_W\leq \frac{d(d+1)}2.
\end{equation}
\end{lem}

\begin{proof}
We need to find the operator coefficients of $\xi_1^r\xi_2^{R-r}$ in the expansion of $(\xi\cdt A)^R$. There are $R+1$ such coefficients.
On the other hand, higher powers $R$ than $d-1$ lead to nothing new, because, by the Cayley-Hamilton Theorem, they can be expressed as linear combinations of lower powers. This leaves at most $\sum_{R=0}^{d-1}(R+1)=d(d+1)/2$ distinct coefficients.
\end{proof}

This lets us expect a high dimensional orthogonal complement of $\MM_W$, In fact, we can give an algebraic description in terms of $A_1$ and $A_2$, of which the commutant is just the simples case.

\begin{prop}\label{prop:incomp} Let $A_1,A_2$ be hermitian matrices of finite dimension, and let $\MM_W$ be
the real linear span of the Weyl ordered monomials. Then, for $P_1,P_2\in\MM_W$
\begin{equation}\label{infnoncomp}
   \tr(A_1P_1A_2P_2)=\tr(A_2P_1A_1P_2).
\end{equation}
In particular, if two density operators $\rho,\rho'$ are related by
\begin{equation}\label{infodiff}
  \rho'=\rho+i(A_1PA_2-A_2PA_1),
\end{equation}
with  $P\in\MM_W$, then they have the same Wigner function.
\end{prop}

\begin{proof}
It suffices to prove \eqref{infnoncomp} for the case $P_1=(A_1+\lambda A_2)^n$, and $P_2=(A_1+\mu A_2)^m$ with real $\lambda,\mu$, which we fix for  the moment. Then for hermitian $X,Y$, consider the real bilinear form
\begin{equation}\label{asyform}
   f(X,Y)=\im\tr(XP_1YP_2).
\end{equation}
This is antisymmetric, because $f(X,X)=\tr\bigl((XP_1X)P_2)\bigr)$ is the trace of a product of hermitian operators, hence real.
We have to show that $f(A_1,A_2)$, i.e., the difference of the two sides in \eqref{infnoncomp}, is zero. Using now the special form of $P_1,P_2$ we get
\begin{eqnarray}
  0 &=& \im\tr\Bigl((A_1+\lambda A_2)^{n+1}\,(A_1+\mu A_2)^{m+1}\Bigr)\\
    &=& f(A_1+\lambda A_2,A_1+\mu A_2) \nonumber\\
    &=& \lambda f(A_2,A_1)+\mu f(A_1,A_2) \nonumber\\
    &=& (\lambda-\mu) f(A_2,A_1). \nonumber
\end{eqnarray}
Hence $f(A_1,A_2)=0$, when $\lambda\neq\mu$. But $f$ depends continuously on these parameters through $P_1$ and $P_2$, so $f(A_1,A_2)=0$ for any choice of $\lambda,\mu$.
\end{proof}

Now experimenting with pairs of random matrices one always finds that the dimension estimate \eqref{dimMMW} is tight, and that \Qref{prop:incomp} gives a complete description of $\MM_W$ and its orthogonal complement. This shows that there are no further algebraic constraints, and suggests that the exceptions (e.g., reducible pairs) are of measure zero. In the same way, a study of triples of matrices suggests that informational completeness almost always holds.

For proving informational completeness, e.g., for the example of spins, we have the following criterion.
\begin{lem}\label{lem:infocLie}
Let $A_1,\ldots,A_n$ be the generators of an irreducible representation of a compact connected Lie group.
Then $\wig$ is informationally complete.
\end{lem}

\begin{proof}
By ``the generators'' we mean that there is a basis of the Lie algebra of $n$ elements, and the $A_k$ are the corresponding generators of one-dimensional subgroups in the given representation.
The exponential operators $\exp(i\xi\cdt A)$ are then just the unitaries of the group representation, written out using the exponential map from the Lie algebra to the Lie group. It is a basic fact that this map is onto \cite{Taoblog2}. Therefore, the product of any two such operators can be written again in the same form. But then the linear span of the operators $\exp(i\xi\cdt A)$ is the same as the span as an operator algebra. This algebra is also clearly closed under adjoints. Therefore, by von Neumann's bicommutant theorem, it is dense if and only if its commutant, the set of operators commuting with all these operators consists of the multiples of the identity. But this is just the assumption of irreducibility of the representation.
\end{proof}

\subsection{Completeness of normally ordered moments}
Other ordering schemes for monomials have been proposed, and each of them will produce its own variant of Wigner functions \cite{CohenBook2}. Let us briefly discuss the informational completeness question for these. ``Normal'' ordering gives the monomials $A_1^{r_1}\cdots A_n^{r_n}$. The Wigner function then has support on the eigenvalue tuples in $\Rl^n$, and (in the non-degenerate case) the corresponding operator coefficients will be rank one operators $\ketbra\phi\psi$, with $\phi$ and eigenvector of $A_1$ and $psi$ and eigenvector of $A_n$. The marginal property holds only for the given operators, but not for their linear combinations. The corresponding ``quasi-probabilities'' are usually complex, so one might want to take the real part \cite{Barut}.
Since the eigenvectors of $A_2,\ldots,A_{n-1}$ hardly enter the game, the most interesting case is $n=2$.

\begin{prop}
Let $A_1,A_2$ be hermitian matrices with non-degenerate spectra. Then the normally ordered moments are complete iff no eigenvector of $A_1$ is orthogonal to an eigenvector of $A_2$.
\end{prop}

\begin{proof}
For any polynomial functions $f,g$,
\begin{equation}\label{stupidmoment}
   f(A_1)g(A_2)=\sum_{ij}f(a_i)g(b_j)\kettbra{\phi_i}\,\kettbra{\psi_j},
\end{equation}
where $a_i,\phi_i$ and $b_j,\psi_j$ are the eigenvalues and eigenvectors of $A_1$ and $A_2$, respectively.
Hence all such expressions are in the span of the normally ordered moments. Taking $f$ to be $1$ on one eigenvalue of $A_1$ and $0$ on the others, and making a similar choice for $g$ shows that the span of the normally ordered moments is exactly the span of the operators $\ket{\phi_i}\braket{\phi_i}{\psi_j}\bra{\psi_j}$. Since without the scalar product factor the $\ketbra{\phi_i}{\psi_j}$ clearly form an orthonormal basis of the Hilbert-Schmidt class, the necessary and sufficient condition for the operators with those factors to be complete is that no $\braket{\phi_i}{\psi_j}$ vanishes.
\end{proof}

Of course, normal (``Wick''-)ordering plays an important role in field theory, when it is applied to creation and annihilation operators to control diverging vacuum expectations. In its application to arbitrary operators \cite{Barut,CohenBook2}, it seems a bit silly.

\subsection{Connection to the BMV-conjecture}
\begin{prop}\label{prop:BMV}
Let $\xi,\eta\in\Rl^n$, and $\xi_0,\eta_0\in\Rl$ such that $(\xi\cdt A+\xi_0\idty)\geq0$ and $(\eta\cdt A+\eta_0\idty)\geq0$, and take $\rho=\idty$.
Then for all $n,m\in\Nl$,
\begin{equation}\label{mixmom}
      \int da\, \wig_\rho(a)\ (\xi\cdt a+\xi_0)^n(\eta\cdt a+\eta_0)^m\geq0
\end{equation}
\end{prop}

Note that with a single factor this just follows from the marginal property. Moreover, each factor is positive on the support of $\wig_\rho$ by \Qref{prop:supp}. Moreover, by changing the basis in the state of $A_k$, maybe shifting by multiples of $\idty$ and integrating out all but the first two variables, we can equivalently state this property for mixed moments in the case $n=2$, with $(\xi\cdt A+\xi_0\idty)=A_1\geq0$, and
$(\eta\cdt A+\eta_0\idty)=A_2\geq0$. Then it reduces via \eqref{WigDefmom} to the statement that all Weyl-ordered moments of two positive operators have positive trace.

This is exactly the equivalent reformulation by  Lieb and Seiringer \cite{LiebBMV} of the Bessis-Moussa-Villani (BMV) conjecture \cite{BMV}. This conjecture says that it is impossible to tell from the functional dependence of a partition function on a linear coupling parameter in the Hamiltonian, whether a system is classical or quantum. More formally, for any Hermitian matrices $A,B$, there is a probability space $(\Omega,\mu)$ and random variables $a,b$ such that
\begin{equation}\label{BMV}
  \tr e^{A+\lambda B}=\int_\Omega\mskip-2mu\mu(d\omega)\, e^{a(\omega)+\lambda b(\omega)}.
\end{equation}
The conjecture remained open for 38 years and was recently proved by H.~Stahl \cite{BMVproof,BMVproof2}, which hence also proves \Qref{prop:BMV}.

We note that the extension to three positive operators fails even in the simplest case. We take the matrices $A,B,C$ to be one-dimensional projections in $2$ dimensions, onto vectors $\phi_1,\phi_2,\phi_3$, say. The coefficient of $\alpha\beta\gamma$ in $\tr(\alpha A+\beta B+\gamma C)^3$ is, up to positive constants,
\begin{equation}\label{BMV3fails}
  \tr(ABC+CBA)=2\re\Bigl(\braket{\phi_1}{\phi_2}\braket{\phi_2}{\phi_3}\braket{\phi_3}{\phi_1}\Bigr).
\end{equation}
Then with $\phi_k=\cos(2\pi k/3)\ket1+\sin(2\pi k/3)\ket2$ each of these scalar products is $\cos(2\pi/3)=-1/2$, so the above product is negative. Note that this also disproves the statement about mixed moments for states other than $\idty/2$, by taking $\rho=C$.

\subsection{Symmetry}\label{sec:covariant}

Since the definition of the Wigner function contains no arbitrary choices, it is automatically covariant with respect to any symmetry respecting the basic setup.
More formally, a {\it symmetry} of an operator tuple $(A_1,\ldots,A_n)$ is a unitary or antiunitary operator $U$ on $\HH$ with the property that
\begin{equation}\label{symultiplet}
  U^*A_kU= \sum_\ell R_{k\ell} A_\ell\ +\alpha_k\idty,
\end{equation}
for some non-singular real matrix $R_{k\ell}$ and constants $\alpha_k$.
	

\begin{prop}\label{prop:cov}
The Wigner function is covariant with respect to any symmetry satisfying \eqref{symultiplet}, in the sense that
\begin{equation}\label{covv}
  \wig_{U\rho U^*}(a)= \frac1{\det R}\,\wig_\rho\bigl(R^{-1}(a-\alpha)\bigr)
\end{equation}
\end{prop}

The proof is completely straightforward, with due care taken of the complex conjugations in the antiunitary case. In the $(P,Q)$ case the symmetries comprise phase space translations and symplectic linear transformations. With bounded operators the set of symmetries will be much more constrained. Indeed, the iteration of $\mathcal T(x)=Rx+\alpha$ must not give an unbounded sequence.

For a basic family of examples, let us take the symmetry group of a regular star or polygon with $p>2$ vertices. Thus $R$ will be the two-dimensional representation of the dihedral group $G=\dihed p$. This consists of the $p$ rotations by angle $2\pi k/p$, and the $p$ reflections around conjugate axes, one of which we choose to be vertical. The representation $U$ will be $p$-dimensional, acting by permutations of the basis vectors, each of which corresponds to one point of the polygon. We take $\rho=\idty/p$ in order to get a fully symmetric Wigner function.
An example with $p=7$ is shown in \Qref{fig:penta}.

The basic question is now: How many free parameters are left for choosing $A_1$ and $A_2$, and how can we systematically generate such examples?
There is a standard method to find the $R$-multiplets in any representation $V$ of $G$, namely to apply the projection
\begin{equation}\label{character}
  \twirl=\frac{\dim R}{\abs G}\sum_g (\tr R_g)\,V_g.
\end{equation}
We want to use this projection for the representation of $G$ acting on $p\times p$-matrices, i.e., $V_g(X)=U_gXU_g^*$. In order to compute the dimension of the range of $\twirl$ we consider the space of operators as a Hilbert space with Hilbert-Schmidt scalar product $\tr(X^*Y)$, with basis $\ketbra jk$. Then
\begin{eqnarray}\label{trtwirl0}
  \tr\twirl&=& \sum_{jk} \tr\bigl((\ketbra kj)\twirl^{(2)}(\ketbra jk)\bigr)   \nonumber\\
     &=& \frac{\dim R}{\abs G}\sum_{g,jk} (\tr R_g)\,\brAAket j{U_g}j\, \brAAket k{U_g^*}k \nonumber\\
     &=& \frac{\dim R}{\abs G}\sum_{g} (\tr R_g)\,\bigl|\tr U_g\bigr|^2.
\end{eqnarray}
Now for the two-dimensional representation $\tr R_g=0$ for any reflection, and for the $p$-dimensional one $\tr U_g=0$ for any non-trivial rotation, because no corner of the $p$-gon is left fixed. Thus the only term contributing to the above sum is the $g=e$, the neutral element. Hence, with $\tr R_e=\dim R=2$,
$\tr U_e=p$, and $\abs G=2p$, we get
\begin{equation}\label{trtwirl}
  \tr\twirl=2p.
\end{equation}
We have one more condition to satisfy, namely that the $2$-direction is a reflection axis of the diagram, so $A_2$ must be invariant under the corresponding reflection. This leaves a $p$-dimensional family of operators $A_2$, from which $A_1$ is obtained as a linear combination of $A_2$ and some $UA_2U^*$.

\section{Examples}
\begin{figure}[ht]
	\includegraphics[width=0.85\columnwidth]{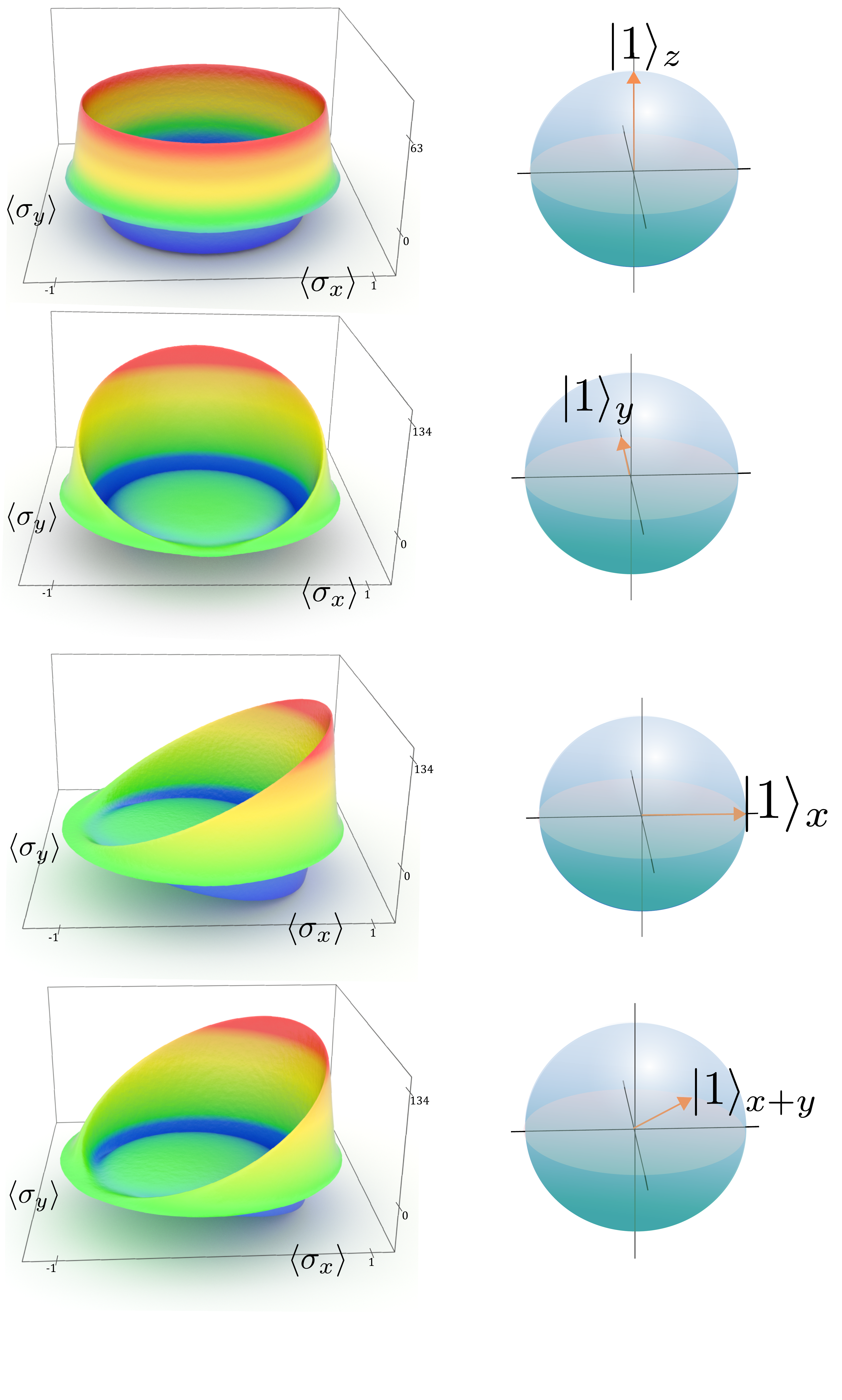}
	\caption{Wigner functions for the pair $(\sigma_x,\sigma_y)$, depicted for the respective spin-up eigenstate of $\sigma_z$, $\sigma_x$, $\sigma_y$ and $1/\sqrt{2}(\sigma_x+\sigma_y)$. Rotations around the vertical axis of the Wigner function and the Bloch sphere match exactly in accordance with \Qref{prop:cov}. }
	\label{fig:qubits}
\end{figure}

\subsection{Qubits, $n=3$}
We take $A_k=\sigma_k$, $k=1,2,3$ to be the Pauli matrices. Differentiating gives
\begin{equation}\label{dqubit}
  \frac\partial{\partial\xi_k}\widehat\wig_\idty= i\tr A_k e^{i\xi\cdt A}.
\end{equation}
Hence, if $\rho=(\idty+\sum_kr_k\sigma_k)/2$ is a general density matrix with $r$ in the Bloch ball, we have
\begin{equation}\label{wigqubit}
  \wig_\rho(a)=\frac12\Bigl(1 + r\cdt a\Bigr)\wig_\idty(a).
\end{equation}
That is, $\wig_\rho$ is obtained from $\wig_\idty$ by multiplication with a positive function. This is true generally whenever $\rho$ is a linear combination of the $A_k$ and $\idty$.

Now $\widehat\wig_\idty(\xi)=2\cos\abs\xi$. The Fourier transform can be calculated explicitly even with a Gaussian factor $\exp(-\veps\xi^2)$. It is the radial function, say $(\wig_\idty*G_\veps)(a)=W_\veps(\abs a)$ with
\begin{eqnarray}\label{wigQBiteps}
  W_\veps(s)&=&\frac1{8 \pi ^{3/2} s\epsilon ^{3/2}}
               \Bigl(e^{-\frac{(s+1)^2}{4 \epsilon }}(s+1){+} e^{-\frac{(s-1)^2}{4 \epsilon }}(s-1)\Bigr) \nonumber\\
      &\rightarrow&\frac{-1}{\pi s}\, \frac d{ds}\frac{e^{-\frac{(s-1)^2}{4 \epsilon }}}{\sqrt{}4\pi\veps}
                     \rightarrow\frac{-1}{2\pi s}\delta'(s-1).
\end{eqnarray}
Here the first simplification in line \eqref{wigQBiteps} is that the first term in the previous equation vanishes exponentially as $\veps\to0$ for all positive $r$, so does not contribute to the limit. The second step is to identify the second term as the derivative of the Gaussian representation of the $\delta$-function.
To check normalization, consider a radial function $f$, which integrates to
\begin{eqnarray}\label{deltaprime}
  \int d^3a\ \wig_\idty(a) f(\abs a)
         &=& 4\pi\int\!\! r^2\,dr \frac{-1}{2\pi r}\delta'(r-1) f(r) \nonumber\\
         &=& 2 \bigl(f(1)+f'(1)\bigr).
\end{eqnarray}
This is correct both for $f=1$, where we get $2=\tr\idty$, and for $f(r)=r^2$, where we get $6=\tr\sum_iA_i^2$. However, one has to be aware that due to negativity \eqref{deltaprime} is not very much like the expectation $\langle f\rangle$. For example, we have
\begin{equation}\label{exparadox}
  \langle(r-1)^2\rangle=0=\langle(r-3)^2\rangle.
\end{equation}
Normally we would conclude from the first equation that the distribution is concentrated on $r=1$ (which is actually true), contradicting the second, which claims with the same right that $r=3$. This is possible because the variance of this distribution is $\langle r^2\rangle-\langle r\rangle^2=3-4=-1$.

To summarize \eqref{wigqubit} and \eqref{wigQBiteps}, for  $\rho=(\idty+\sum_kr_k\sigma_k)/2$ with Bloch ball coordinate $r\in\Rl^3$, $\abs r\leq1$, we have
\begin{equation}\label{wigqubit2}
  \wig_\rho(a)=\frac{1 + r\cdt a}{4\pi \abs a}\, \delta'(1-\abs a).
\end{equation}

Note that $\wig_\rho$ vanishes exactly in the interior of the Bloch ball. This is no longer true if we integrate out one of the variables.

\subsection{Qubits, $n=2$}\label{sec:qbitn=2}

Apart from the standard $(P,Q)$-case this is apparently the only example for which Wigner functions in our sense have been computed \cite{CohScu}. However, the result as reported by Cohen and Scully \cite{CohScu} makes no sense, since it contains a $\delta$-distribution multiplied with a function that is infinite precisely on the line where the $\delta$-function is non-zero. Therefore, we redo the computation, using the same basic trick.  We confirm the result \cite{CohScu} for the regular part, but give a better explanation of how to understand the singularity of this distribution.

Exactly as for $n=3$, the general case is reduced to the case $\rho=\idty/2$, so the analogue of \eqref{wigqubit} holds, when the component $r_3$ is set equal to zero. The resulting state dependence is shown in \Qref{fig:qubits}.

By rotation invariance it suffices to consider an $\veps$ regularized version at $a=(r,0)$ and evaluate the $\xi$-integral in polar coordinates $(s,\phi)$:
\begin{eqnarray}
  \wig_{\idty/2,\veps}(r,0)&=& \frac1{(2\pi)^2}\int s\,ds\,d\phi\  e^{-irs\cos\phi-\veps s} \cos s \nonumber\\
     &=& \frac1{2\pi} \int_0^\infty ds\ s\cos(s) e^{-\veps s}\ J_0(rs)\nonumber\\
     &=& \frac1{2\pi}\left.\frac d{d\lambda}\int_0^\infty ds\ \sin(\lambda s) e^{-\veps s}\ J_0(rs) \right|_{\lambda=1}\nonumber\\
     &=& \frac1{2\pi}\left.\frac d{d\lambda}\im\frac1{\sqrt{(\veps-i\lambda)^2+r^2}} \right|_{\lambda=1}
\end{eqnarray}

\begin{figure}[t]
	\includegraphics[width=0.85\columnwidth]{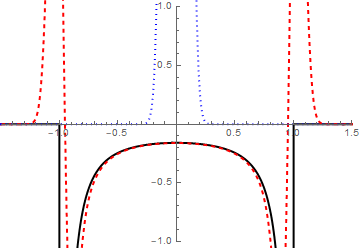}
	\caption{Section ($y=0$) through the Wigner function $\wig_{\idty/2}$ for $(\sigma_x,\sigma_y)$ or, equivalently, the case $\ket1_z$ in \Qref{fig:qubits}.
		Black solid line: the exact regular part \eqref{qubit2b}. Red, dashed: numerical calculation of the the whole distribution regularized with a Gaussian shown
		in the center (dotted). }
	\label{fig:qubitsect}
\end{figure}

The integral over the Bessel function at the third equality follows from \cite{gradst}[6.611.1]. After integrating this with a test function $f$ we can take the limit $\veps\to0$, so
\begin{eqnarray}\label{qubit2int}
  \int\!da\  \wig_{\idty/2}(a)f(a)
                    &=& \frac1{2\pi}\left.\frac d{d\lambda}\int\!\!da\ f(a)\im\frac{-i}{\sqrt{\lambda^2-\abs a^2}}\right|_{\lambda=1}\nonumber\\
                    &=& \frac1{2\pi}\left.\frac d{d\lambda} \int_{\abs a<\lambda}\!\!da\ \frac{f(a)}{\sqrt{\lambda^2-\abs a^2}}\right|_{\lambda=1}\nonumber\\
                    &=& \frac1{2\pi}\left.\frac d{d\lambda}\int_{\abs b<1}\!\!db\ \frac{\lambda f(\lambda b)}{\sqrt{1-\abs a^2}}\right|_{\lambda=1}\nonumber\\
                    &=& \frac1{2\pi}\int_{\abs a<1}\!\!da\ \frac{f(a) + a\cdot \nabla f(a)}{\sqrt{1-\abs a^2}}
\end{eqnarray}
Here, at the third equality we substituted $b=\lambda a$, $db=\lambda^2da$. The integral \eqref{qubit2int} makes sense as written, because $f$ is a test function, hence differentiable, and $(1-a^2)^{-1/2}$ is an integrable function. For writing it just as an integral over $f$, not its derivatives, we use partial integration, which holds by definition of the distributional derivative:
\begin{eqnarray}\label{qubit2a}
   \wig_{\idty/2}(a)&=& \frac1{2\pi}\left(\frac1{\sqrt{1-\abs a^2}}-\sum_j \frac\partial{\partial a_j}\frac{a_j}{\sqrt{1-\abs a^2}}\right)\ \\
          &=& \frac{-1}{2\pi(1-\abs a^2)^{3/2}} +\mbox{boundary}. \label{qubit2b}
\end{eqnarray}
Here \eqref{qubit2a} is correct as a mere rewriting of \eqref{qubit2int}, provided the derivative is considered as a distributional derivative, and the function is continued as $\wig(a)=0$ for $\abs a>1$.
Equation \eqref{qubit2b} is obtained by evaluating the derivatives in the sense of ordinary functions. That is correct even without the boundary term for the integral with any test function with support strictly inside the unit circle. So this expression describes the ``non-singular part'' of the Wigner function, and is uniquely determined by this property.

As an expression for the whole distribution, however, \eqref{qubit2b} is misleading, because it suggests that $\wig_{\idty/2}(a)$ can be written as the sum of the non-singular part and some purely singular part supported on the circle. This is not true, because the regular part by itself is not integrable, so not even a distribution. So any attempt to use \eqref{qubit2b} term by term results in an indeterminate expression.
\begin{figure*}[t]
	\label{fig:spin4}
	\includegraphics[width=0.99\textwidth]{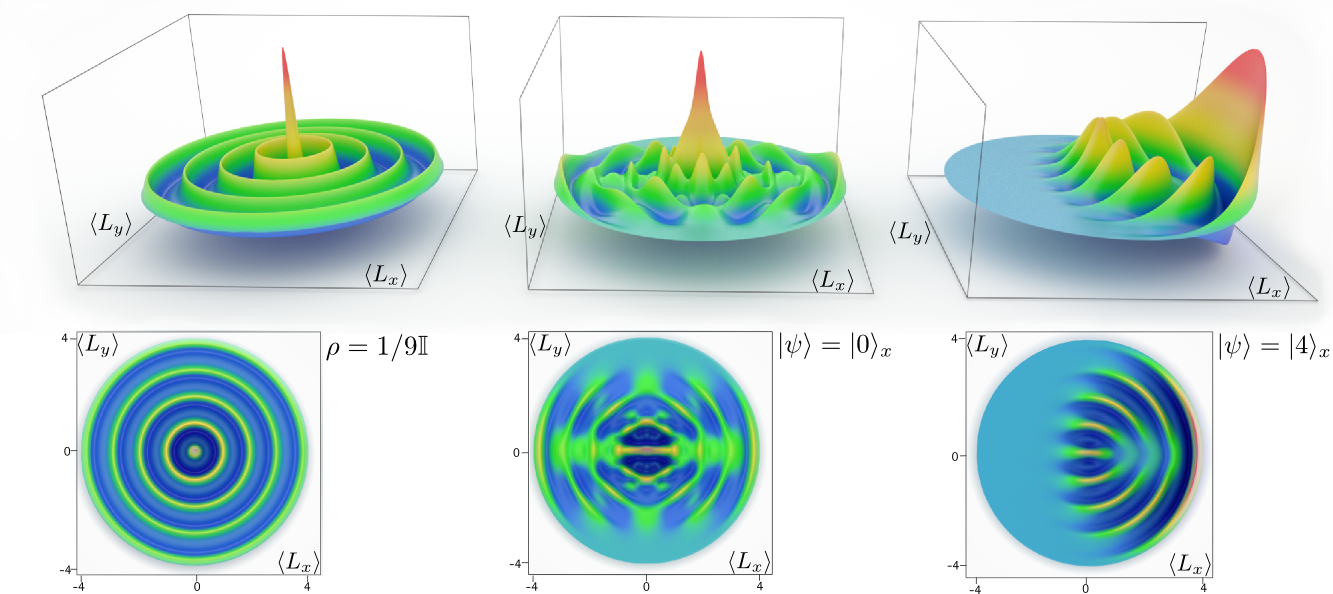}
	\caption{\coloronline Wigner function of angular momentum components $L_x$ and $L_y$ for spin $s=4$, depicted for the maximally mixed state, the $j_x=0$ and the $j_x=4$ eigenstate. The Wigner fuction of the maximally mixed state shares the full rotational symmetry. The Wigner function of the $j_x=0$ state has only reflection symmetries on the $x=0$ and the $y=0$ axis.  The Wigner function of the $j_x=4$ state is clearly concentrated in the $\langle L_x \rangle \geq 0$ region and has only a symmetry on the $y=0$ axis.}
\end{figure*}
\subsection{Spins}
The angular momentum operators are one of those case in which all linear combinations are equally relevant physical operators. It is this case which originally motivated us \cite[Prop.\,7]{AMU}. In the spin-$s$ representation of SU(2), the singular manifold is the collection of concentric spheres of radius $\hbar m$. Moreover, the operators $\exp(i\xi\cdt A)$ are just the unitaries of the group representation. Hence their linear span is an algebra, whose closure is dense by irreducibility of the representation. Hence $\widehat\wig_\rho$ and consequently $\wig_\rho$ determines $\rho$ uniquely. Further properties, including the classical limit, will be analysed in a future publication.

\section*{Conclusion and Outlook}
The Wigner distributions considered in this paper at first sight look very  different from  the usual ones (singular distributions vs bounded continuous functions). Since the definition is the same, just applied to different operators, this points to some special features of the case of canonical operators, to which we have  become accustomed, and which make the case of canonical pairs special. The high phase space symmetry is certainly a key, but also the fact that all spectra are absolutely continuous, and hence less prone to producing singularities. Of course, in this initial exploration we focused on the singularities as the most conspicuous feature. But it would be interesting to consider intermediate cases.

In this regard it would be interesting to see if some of the group-based distribution functions can be retrieved by integration. For example, for an angle variable $\phi$ one can avoid artificial branch cuts by considering the two hermitian operators $(\cos\phi,\sin\phi)$ (e.g., \cite{Kastrup}). The joint distributions of this pair with a momentum or number variable would probably produce some distribution away from the unit circle, but a radial integration in the plane would eliminate that, while retaining the overall symmetry of the problem.

Another question is posed by the simple form in the qubit case: Can $\wig_\rho$ maybe be written as a non-singular factor times the distribution $\wig_{\idty}$, which depends only on the observables chosen? This separation of a factor coming from the observables and a factor coming from the state, even if the factors were combined in a slightly more complicated way, would be very useful. A related open question is the development of simple inversion formulas, for example in the case of spins.

The Wigner distribution is an object canonically associated to any tuple of operators and a state, and this already makes it worthwhile to find out what story it has to tell. We have shown that there is a remarkable richness in the structure of singularities. But many more details will have to be explored to make it a tool that speaks more directly to intuition. Among these will be a detailed study of the case of angular momentum including the classical limit, and the transition to the Holstein-Primakov approximation. In that approximation we get a phase space with standard position and momentum operators, but also the oscillator Hamiltonian as a third operator. This is an enhancement of the standard Wigner function of some interest in its own right: It resolves the paradox that while the energy function on phase space takes continuous values, its Wigner quantization has discrete spectrum.

Further worthwhile topics are (1) the analysis of the structure of the singularities as in Sect.~\ref{sec:qbitn=2}, but for generic singular lines using perturbation theory. (2) A rigorous analysis including the unbounded case, under the natural condition that the exponential series for all $\exp(i\xi\cdt A)$ make sense on a common dense domain. (3) A derivation of uncertainty bounds of measurement or preparation type \cite{AMU,MUR} between the $A_k$ in terms of the Wigner function. (4) The investigation of infinite dimensional observables.

\newpage
\section*{Acknowledgements}
We acknowledge support from the RTG 1991 and CRC 1227 DQ-mat founded by the DFG, and the BMBF project Q.Link.X.
R.S. gratefully acknowledges A. Ketterer and A. Asadian for inspiring discussions.
\bibliography{Wigref}
\end{document}